\documentstyle[amssymb,stmaryrd,epsfig]{mnv2}
\title[The stellar populations of isolated galaxies]
{The spatially resolved stellar populations of isolated early-type
  galaxies}

\author[F.M. Reda {\it et al.\/}]{Fatma M.
Reda$^{1}$\thanks{freda@astro.swin.edu.au,rproctor@astro.win.edu.au,
dforbes@swin.edu.au,george.hau@durham.ac.uk, slarsen@eso.org},
Robert N. Proctor$^{1}$, Duncan A. Forbes$^{1}$, George K.T. Hau$^{2}$, \\
\LARGE
S{\o}ren S. Larsen$^{3}$\\
  $^1$Centre for Astrophysics \& Supercomputing, Swinburne University, 
Hawthorn, VIC 3122, Australia\\
$^2$Department of Physics, University of Durham, South Road, Durham, 
DH1 3LE, UK\\
  $^3$Astronomical Institute, University of Utrecht, Princetonplein 5,
 3584 CC Utrecht, The Netherlands\\
}
               
\begin{document} 

\date{Accepted ***. Received ***}

\pagerange{\pageref{firstpage}--\pageref{lastpage}} \pubyear{***}

\maketitle  

\begin{abstract}
We present radial stellar population parameters for a subsample of 12
galaxies from the 36 isolated early-type galaxies of Reda et al.   
Using new long-slit spectra, central values and radial gradients for
the stellar age, metallicity [Z/H] and $\alpha$-element abundance [E/Fe]
are measured. Similarly, the central stellar population parameters are
derived for a further 5 isolated early-type galaxies using
their Lick indices from the literature.
On average, the seventeen isolated galaxies have mean central
[Z/H]$_o$ and [E/Fe]$_o$ of $0.29\pm0.03$ and $0.17\pm0.03$
respectively and span a wide range of ages from 1.7 to 15 Gyrs. 
We find that isolated galaxies follow similar scaling relations between
central stellar population parameters and galaxy velocity
dispersion to their counterparts in high density environments. However,
we note a tendency for isolated galaxies to have slightly younger ages,
higher metallicities and lower abundance ratios. Such properties are
qualitatively consistent with the expectation of an extended star
formation history for galaxies in lower density environments.
Generally we measure constant age and [E/Fe] radial gradients. 
However, three galaxies show remarkable positive age 
gradients and two galaxies have negative age gradients. We find
that the age gradients anti-correlate with the central galaxy
age. Thus as a young starburst evolves, the age gradient flattens 
from positive to almost zero.
Metallicity gradients range from near zero to strongly negative. For
our high mass galaxies ($\sigma > 160$ km/s) metallicity gradients 
are shallower with increasing mass. Such behaviour is not predicted in
dissipational collapse models but might be expected in multiple
mergers. 
The metallicity gradients are also found to
be correlated with the central age and metallicity, as well as to the
age gradients. In conclusion, our stellar population data for a sample
of isolated 
early-type galaxies are  more compatible with an extended
merger/accretion history than early dissipative collapse.
\end{abstract}


\section{Introduction}

The stellar population properties of early-type galaxies
(e.g. age, metallicity and $\alpha$-element abundance [E/Fe])
provide crucial clues to their evolutionary history. Most work to date
has concentrated on galaxy central 
regions. While useful, such data only sample a small fraction of
the galaxy mass and do not provide any radial gradient information. 

Radial gradients can help discriminate between different
formation models. For example, dissipational collapse models
(Larson 1974; Carlberg 1984; Kawata \& Gibson 2003) predict
strong metallicity gradients that correlate with galaxy
mass, whereas mergers tend to result in shallow gradients (White
1980; Bekki \& Shioya 1999) with little galaxy mass dependence. As [E/Fe] may
be an indicator of 
star formation timescale, a positive gradient indicates
outside-in formation and a negative one the opposite (e.g. Ferreras \&
Silk 2002). Age gradients indicate whether any young 
stars are confined to the central regions, and hence an indication of
their mass contribution. 

Kobayashi \& Arimoto (1999) compiled a list of 80 early-type
galaxies with radial metallicity gradient measurements; the average value
being --0.3 dex per dex.
For Coma cluster ellipticals, Mehlert et al. (2003) confirmed
significant radial metallicity gradients but, on
average, found no radial age or [E/Fe] gradients. In contrast,
S{\'a}nchez-Bl{\'a}zquez et al. (2006a) found significant age gradients in
a sample of cluster, group and field early-type galaxies. A weak
correlation between metallicity gradient and galaxy mass was
found by S{\'a}nchez-Bl{\'a}zquez et al. (2007) and Forbes,
S{\'a}nchez-Bl{\'a}zquez \& Proctor (2005) for cluster ellipticals.

The number of studies that have focused on the stellar
populations of early-type galaxies in very low density
environments is small. Kuntschner et al. (2002) obtained spectra
of three E and six S0 galaxies in low density environments
(i.e. they had less than two bright neighbours within 1.3 Mpc and
$\pm$350 km/s). Five of their nine galaxies revealed emission
indicating ongoing star formation. Using Lick absorption lines
and Vazdekis (1999) single stellar population (SSP) models they found
their galaxies to be on average younger and more metal-rich than
cluster galaxies. However, they noted the dominance of S0s in their
sample. They also found a similar [E/Fe] vs velocity dispersion
relation as for cluster galaxies.  Collobert et al. (2006)
extended the Kuntschner et al. work with spectra of 22 galaxies,
for Hubble types ranging from S0/a to pure E. Half came from the 2dFGRS
survey (with no bright companions within 1 Mpc) and half from the
low density sample of Colbert et al. (2001) (no RC3 catalogued
galaxies within 1 Mpc and $\pm$1000 km/s). After applying
emission line corrections to 7 of their 22 galaxies, they applied
Thomas, Maraston \& Bender (2003) models to 7 Lick absorption lines
with a $\chi^2$ fitting 
technique. They found a spread to younger ages than for cluster
ellipticals, but no clear [E/Fe] vs velocity dispersion
relation. They speculated that isolated ellipticals {\it
assembled} quite recently. Both studies only probed the central
regions of their sample galaxies.

Here we extend the previous work by examining the radial stellar
population properties to around 1 effective radius in a sample of a 12
isolated galaxies. The sample 
presented here comes from the well-defined and highly isolated
sample of early-type galaxies from Reda et al. (2004). 
Our new data are supplemented  by data for 5 isolated galaxies from
the literature.
We also utilize the latest SSP models and fitting methods.

\section{The data}
In Reda et al. (2004), we defined a sample of 36 isolated early-type
galaxies in the local universe ($z < 0.03$). Here we present new data
on the stellar populations of a subsample for 12 galaxies from that
sample. The basic data of these galaxies and their sources are
summarised in Table 1.
We supplement our data with data from the literature for other isolated
galaxies of Reda et al. (2004).

Denicol{\'o} et al. (2005a) extracted 21 Lick
spectral indices  for the central r$_e/8$ region for a  sample of 86
early-type galaxies. 
Six of our 36 isolated galaxies were included in their
study. Denicol{\'o} et al. applied an emission correction to the indices
where the galaxy spectra show evidence of emission lines. Comparing their
extracted stellar population parameters to the literature, their
measurements tend to be younger and more metal-rich (Denicol{\'o} et
al. 2005b). 
Using their published indices of the six isolated galaxies, and
applying the multi-index $\chi^2$ minimization technique which we are
using for our observations and the same SSP model (see Sec. 2.5), we
have extracted the  
central stellar population parameters for these galaxies which are
listed in Table 2. 

The spatially resolved stellar population of the isolated galaxy NCG
821 was previously studied by Proctor et al. (2005). Here we used their
data to extract the central values of the age, total metallicity [Z/H] 
and $\alpha$-elements abundance [E/Fe] within  r$_e/8$ which are also
listed in Table 2. These measurements of the central
parameters are consistent to our measurements using the indices from
Denicol{\'o} et al. (2005a). 
 We also measured the radial gradient of
these parameters, considering all apertures beyond the seeing limit, to
be 0.29$\pm$0.05, $-0.06\pm$0.03, $-0.72\pm$0.04 for log(age), [E/Fe]
and [Z/H] respectively.

\begin{table*}
 \begin{minipage}{170mm}
\begin{center}
\begin{tabular}{llcllrll}
\multicolumn{8}{l}{\bf Table 1. \small Basic data for the observed isolated galaxies.}\\
\hline
 Galaxy       & Dist.& $M_K$  & $M_B$   & $M_B$   &$\log r_e$& $\pm$ & $\log r_e$\\
              & (Mpc)& (mag)  & (mag)   & source  & (pc)     &       & source    \\
\hline		        	 	 	            	 
NGC  682      &  73 & --24.88 & --19.9  & Paper I  & 3.73 &  0.10 & PS97         \\
NGC 1045      &  60 & --25.02 & --20.9  & Paper I  & 3.66 &  0.10 & Paper I      \\ 
NGC 1162      &  51 & --24.53 & --20.7  & Paper I  & 3.63 &  0.10 & 2MASS        \\ 
NGC 2271      &  32 & --23.84 & --20.0  & Paper II & 3.47 &  0.07 & Paper II     \\ 
NGC 2865      &  35 & --24.26 & --20.7  & Paper I  & 3.68 &  0.01 & Paper I      \\  
NGC 4240      &  26 & --22.71 & --18.8  & Paper I  & 3.29 &  0.10 & 2MASS        \\  
ESO 153-G003  &  84 & --25.04 & --20.9  & Paper II & 3.62 &  0.06 & Paper II     \\  
ESO 218-G002  &  54 & --24.51 & --20.9  & Paper I  & 3.76 &  0.11 & Paper I      \\ 
ESO 318-G021  &  62 & --24.43 & --20.7  & Paper II & 3.77 &  0.01 & Paper II     \\ 
MCG-01-27-013 & 121 & --25.08 & --21.4  & Paper I  & 3.98 &  0.05 & Paper I      \\ 
MCG-02-13-009 &  73 & --24.58 & --21.3  & Paper I  & 3.67 &  0.10 & 2MASS        \\  
MCG-03-26-030 & 119 & --25.79 & --21.7  & Paper I  & 3.92 &  0.02 & Paper I      \\ 
\hline 
\end{tabular}
\end{center}
Note: Distances are obtained using the Virgo corrected 
recession velocities (from LEDA) with H$_0$ = 75 km s$^{-1}$
Mpc$^{-1}$. $M_K$ and $M_B$ are the absolute magnitudes in the $K$ and
$B$-bands respectively. $M_K$ are from 2MASS catalogue and $M_B$ from
Reda et al. (2004, Paper I) and Reda et al. (2005, Paper
II). Effective radius 
r$_e$ from the 2MASS catalogue is the isophotal radius $r_{20}$ in the
$K$-band  (via LEDA database) converted to effective radius using the
correlation in Jarrett et al. (2003, Fig. 11). For NGC 682 the
effective radius is derived from the fundamental plane (Prugniel \& Simien
1997; PS97)
\end{minipage}
\label{basic}
\end{table*}

\begin{table*}
\begin{minipage}{170mm}
\begin{center}
\begin{tabular}{lccccc}
\multicolumn{5}{l}{}\\

\multicolumn{5}{l}{\bf Table 2. \small Central (r$<$r$_e/8$) stellar
  populations of isolated  galaxies from literature.} \\
\hline
Galaxy  & log($\sigma_o$) (km/s)& age$_o$ (Gyr) &  [E/Fe]$_o$  & [Z/H]$_o$  \\
 \hline  
NGC 821 & 2.33$\pm$ 0.06 &  4.7 $\pm$ 1.9 & 0.15$\pm$ 0.02 &   0.48$\pm$ 0.05 \\ 
NGC 1045& 2.36$\pm$ 0.06 &  5.3 $\pm$ 1.4 & 0.12$\pm$ 0.02 &   0.43$\pm$ 0.03 \\
NGC 1132& 2.37$\pm$ 0.05 & 11.2 $\pm$ 4.7 & 0.36$\pm$ 0.04 &   0.10$\pm$ 0.07 \\ 
NGC 2128& 2.25$\pm$ 0.06 &  2.8 $\pm$ 1.2 & 0.27$\pm$ 0.03 &   0.48$\pm$ 0.09 \\ 
NGC 6172& 2.13$\pm$ 0.05 &  2.5 $\pm$ 0.5 & 0.00$\pm$ 0.03 &   0.13$\pm$ 0.04 \\ 
NGC 6411& 2.25$\pm$ 0.03 & 10.6 $\pm$ 0.8 & 0.18$\pm$ 0.04 & --0.05$\pm$ 0.02 \\ 
\hline 
NGC 821 & 2.29$\pm$ 0.01 &  5.6 $\pm$ 0.5 & 0.22$\pm$0.01  & 0.41$\pm$  0.06 \\
\hline 
\end{tabular}
\end{center}
Notes: Parameters of the first six galaxies are obtained using the
published 
Lick indices of Denicol{\'o} et al. (2005a) and refit to a SSP model for
this work. For NGC 821, the parameters are 
the average of all apertures within r$<$r$_e/8$ from Proctor et al. (2005).
\end{minipage}
\label{centr1}
\end{table*}

\subsection{Observations}
Our spectroscopic observations were carried out using EFOSC2
at the  ESO 3.6m telescope on  the La Silla Observatory in two
observing  runs,  2002 Jan. 16-18  and  2004 Dec.  11-12. 
The observational set
up and slit position are described in detail by Hau \& Forbes
(2006). Lick and spectrophotometric standard stars were observed at
the parallactic angle. The observing nights were
photometric and the seeing was generally better than $1.0''$. 

Observations of the first run (2002) were collected using a slit of
$300''$ length and $1.5''$ width which was positioned along the major
axis of each galaxy, with exception for the S0 galaxy ESO 153-G003
where the slit was positioned along the minor axis to avoid the disc.
Using the ESO grism\#8 of 600 line/mm provides a
spectral resolution of 9.3 \AA~FWHM. The accumulated exposure time was 2
$\times$ 1200 seconds for each galaxy. 
In the 2004 run we used a $300'' \times 1.2''$ slit which provides a spectral
resolution of 7.8 \AA~FWHM. The accumulated exposure time was 3 $\times$
1200 seconds for each galaxy.
The multiple observations of each galaxy are combined to increase the
S/N ratio. 

Additionally, a number of spectrophotometric stars were observed for flux
calibration and Lick standard stars (of type between K0 and K5) to be
used as templates for velocity dispersion measurements and for
calibration of the line-strength indices to the Lick/IDS system.
A number of bias, dark current and dome flat were recorded each night.
Observations of all galaxies and stars gave a wavelength coverage of
4300-6300 \AA. 

\subsection{Basic data reduction}
Data reduction, including bias, dark current and flat field
subtraction, is performed using tasks within IRAF. 
Wavelength calibration is done using Helium-Argon lamp and is good to
within 0.5 \AA, while the flux standard stars are used to
calibrate the spectra.
The 2-D spectra are corrected for the S-distortion and sky subtracted.
For more details about the data reduction refer to
Hau \& Forbes (2006). 

Finally the 2-D spectra of the galaxies are spatially binned using
the {\small STARLINK} task {\small EXTRACT} to obtain 1-D spectra of
S/N ratio greater than 20. We will refer to these 1-D spectra of each
galaxy as {\it apertures}.  

\subsection{Velocity Measurements}
The heliocentric recession velocity of the standard stars are
obtained by comparing their spectra to the solar spectrum. The {\it
fxcor} task within {\small IRAF} is used to fit a
Gaussian to the cross-correlation function to determine the maximum
peak. The position of the maximum peak centre gives the required
heliocentric recession velocity of the stars.

Once the heliocentric recession velocity of the standard stars were measured, 
the redshifts and velocity dispersions of the galaxies are determined by 
comparing their spectra to the spectra of the standard stars.
 The {\it fxcor} task of {\small IRAF} is used to compute the
cross-corelation of the galaxy spectra with the spectra of each 
template star. Once the correlation is computed, the position of the
maximum peak centre gives the recession velocity V$_r$ and
the velocity dispersion $\sigma$. The final V$_r$
and $\sigma$ are the average of the values estimated from all the 
template stars. 
The radial kinematics for the same sample of galaxies are discussed in
Hau \& Forbes (2006).

\subsection{Absorption line-strength measurements}

\begin{table}
\begin{center}
\begin{tabular}{lc|cc|cc|}
\multicolumn{6}{l}{\bf Table 3. \small Offset of the Lick indices from
  the published values.}\\ 
\hline
 & & \multicolumn{2}{|c|}{2002 run} &
     \multicolumn{2}{|c|}{2004 run} \\
\hline
Index & unit &  mean  & error of & mean   & error of \\
      &      & offset & the mean & offset & the mean  \\
\hline 
Fe4383 &  \AA   & --0.031 &  0.329  &  --0.229 &  0.120  \\
Ca4455 &  \AA   &  0.214 &  0.152   &  --0.054 &  0.171  \\
Fe4531 &  \AA   & --0.163 &  0.132  &  --0.342 &  0.103  \\
C4668  &  \AA   & --0.828 &  0.097  &  --0.598 &  0.187  \\ 
H$\beta$ & \AA  & --0.018 &  0.088  &  --0.041 &  0.113  \\
Fe5015 &  \AA   & --0.284 &  0.079  &  --0.357 &  0.185  \\ 
Mg$_1$  &  mag  &  0.013 &  0.002   &    0.019 &  0.007  \\
Mg$_2$  &  mag  &  0.019 &  0.003   &    0.018 &  0.006  \\ 
Mg$_b$  &  \AA  & --0.083 &  0.071  &    0.039 &  0.195  \\  
Fe5270 &  \AA   & --0.135 &  0.053  &  --0.412 &  0.147  \\ 
Fe5335 &  \AA   & --0.214 &  0.093  &  --0.274 &  0.215  \\ 
Fe5406 &  \AA   & --0.121 &  0.035  &  --0.302 &  0.059  \\
Fe5709 &  \AA   & --0.006 &  0.128  &    0.031 &  0.065  \\
Fe5782 &  \AA   &  0.084 &  0.026   &    0.027 &  0.067  \\  
Na$_D$  & \AA   &  0.058 &  0.132   &  --0.239 &  0.138  \\
TiO$_1$ & \AA   &  0.004 &  0.005   &    0.006 &  0.003  \\
\hline
\end{tabular}
\end{center}
\label{offs}
\end{table}

We adopted the definition of Lick/IDS absorption line indices from
Trager et al. (1998). They defined each index using a pair of
pseudocontinua bracketing the line feature and the central passbands
of the line itself.  

Our observations cover a wavelength range of 4300-6300 \AA~which includes 
16 Lick/IDS index. These are one Balmer index (H$\beta$), three
Magnesium indices (Mg$_1$, Mg$_2$ and Mg$_b$), Calcium (Ca4455), Sodium (NaD),
Titanium Oxide (TiO$_1$) indices plus nine Iron indices Fe4383, Fe4531, Fe4668
(referred to as C4668), Fe5015, Fe5270, Fe5335, Fe5406, Fe5709 and Fe5782.

A number of Lick standard stars were observed during each observing
run and will be used in the following sections to calibrated the
measured indices to the Lick system. 

\subsubsection{Matching to the Lick resolution}
Lick absorption indices are affected by a wavelength dependent
instrumental resolution. This resolution ranges from minimum value of
$\sim 8.4$ \AA~in the range 4900- 5400 \AA~and degrade at the blue and red
sides reaching above $\sim 10$ \AA~(see Fig. 7 in Worthey \& Ottaviani
1997).  
Our observed instrumental resolutions (9.3 \AA~and 7.8 \AA~for 2002
and 2004 runs respectively) are different from those of the standard
Lick system. The galaxy velocity dispersion will also alter the
observed resolution of the absorption lines.  

To transform to the standard Lick system, two different methods are
used according to the observed total resolution  of the spectra
(including both instrumental and velocity dispersion effects). 
For indices with combined resolution higher than that of the
Lick system, spectra are degraded to the resolution of the Lick/IDS
library by convolving the spectra with a wavelength dependent Gaussian.
While for indices with resolution lower than the  Lick/IDS system,
a correction factor is measured by convolving the spectra of five Lick 
standard stars  with a series of Gaussians  widths from 0 to 500 km/s. 
Comparing the indices from the
artificially broadened stellar spectra to the un-broadend indices, we 
estimate the index corrections for the broadening effect of velocity 
dispersion in galaxies. This factor is then used to
transform the indices measured in the galaxies of low resolution
spectra to the Lick resolution (Proctor
\& Sansom 2002; Proctor et al. 2004).

\subsubsection{Zero-point offset due to flux calibration}

Furthermore, the original Lick/IDS spectra were not flux calibrated
while our spectra 
are. To compensate for the resulting effect on the shape of the
spectral continuum, the differences between the measured and the
published indices values are obtained for each observed calibration
star. For each index, the values from all stars are averaged and
compared to the literature values to give the mean offset. The error
of the mean is obtained as:  
$\frac{\sigma _{offset}} {\sqrt{N}}$ 
where $\sigma _{offset}$ is the $rms$ of the scatter about offset and
$N$ is the number of the standard stars. The adopted offset of each
index and its error are given in Table 3. 

\subsection{Galaxy stellar population parameters}

\begin{figure}
\centerline{\psfig{figure=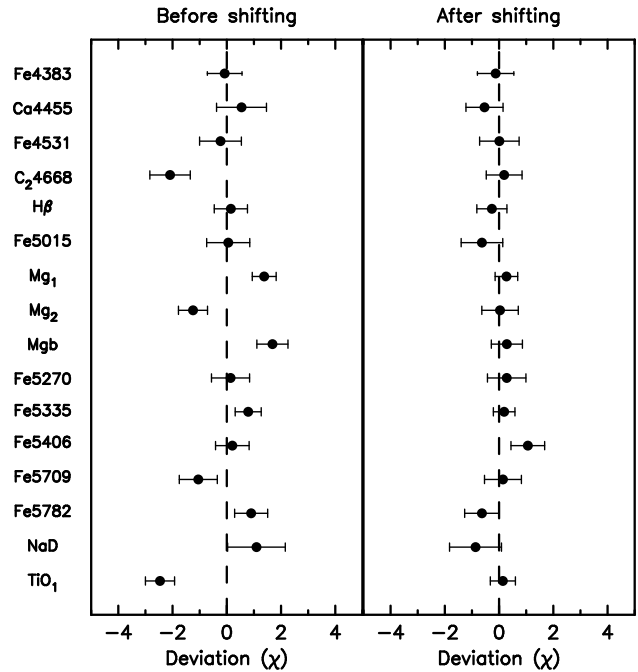,width=0.47\textwidth}}
\caption
{
The average deviation in units of error (i.e. $\chi$) of the 16
indices from the best fit values of NGC 2271 as a typical galaxy of
the sample. Indices of the 31 apertures of the galaxy are shown  before
(left) and after (right) applying the shifting technique (see text for
details). Error bars represent the rms scatter in the deviation. 
}
\label{chi}
\end{figure}

The age, total metallicity [Z/H] and $\alpha$-element abundance
[E/Fe] properties of our sample galaxies are estimated by comparing the Lick
absorption lines indices to the single stellar population (SSP) 
model of Thomas, Maraston \& Korn (2004, TMK04). 
The [E/Fe] parameter measures the ratio of the $\alpha$-elements (N, O,
Mg, Na, Si, Ti) to the Fe-peak elements (Cr, Mn, Fe, Co, Ni, Cu, Zn).

The multi-index $\chi^2$ minimization technique (Proctor \&
Sansom 2002; Proctor et al. 2004) is used to obtain the corresponding
SSP model values of the age, [Z/H] and [E/Fe] abundance ratios for 
the galaxies. In this technique as many indices as possible are fit to
the model. 
Because most indices contain some information regarding
each of the stellar parameters (Proctor \& Sansom 2002), using this
fitting technique has advantages of 
including all possible information recorded in all absorption line
indices to measure these parameters.
Another advantage of this fitting technique is that it identifies
indices that are highly deviant from model values, permitting their
exclusion from the fitting process. In addition, we note that
parameters measured in this way are less prone to
uncertainties in data reduction such as flux calibration, stray cosmic
ray and also less prone to weak emission effects (Proctor, Forbes \&
Beasley 2004). 
Therefore we start by fitting all 16 indices. The indices
which deviate by more 
than $3\sigma$ are clipped and the $\chi^2$ fit recalculated.  
After clipping the highly deviated indices ($> 3\sigma$), we
notice that, for some apertures within each galaxy, the fitting is still
not stable and the estimated values of the age, metallictiy and [E/Fe]
are significantly affected by clipping different indices.  
Also,  after this initial fitting process, many spectral indices show a
significant range of deviations from the best fitting
values. Considering NGC 2271 as a typical galaxy of the sample, the
left panel of Fig. \ref{chi} shows the average deviation in units of error
(i.e. $\chi$) of the 16 indices from the best fit values. 

Furthermore, we notice that some indices lie outside the
model grid which can be due to residual offsets between the measured
indices and the standard system, unknown dependencies of the line
strength predictions on any abundance ratio or inaccuracies in SSP
models. To eliminate the effects of the 
remaining offsets, we implemented the technique described by Kelson et
al. (2006) to shift the zero points of the model to our data. Kelson
et al. defined the reference point as a set of previously published
stellar population parameters for the central 
massive early-type galaxy in the cluster of their study. To apply this
technique to our galaxy sample we define a local 
reference point for each galaxy by choosing two central apertures with
high S/N ratios that show higher stability during the initial 
fitting to the model. 
The initial measured age, [Z/H] and [E/Fe] of these two central
apertures are considered as a reference zero point for that galaxy. 
We measure the offset between each index from the corresponding
values in the model, at fixed age and metallicity as estimated by the
initial fit. Then we apply similar shifts to the indices of all other
apertures. 
After applying this shift, the number of highly deviant indices
from the model is reduced and consequently larger number of indices
are used during the fitting process which become much faster. 
Examining the extracted stellar population parameters before
and after shifting the indices, we find that they are similar. That
means applying Kelson et al. method only reduces the scatter of the
indices around the model.
Fig. \ref{chi} shows the average deviation of the indices before and
after applying Kelson et al. technique. 
Errors in the derived parameters are estimated using Monte Carlo type
realizations of the best-fitting SSP models perturbed by the index
error estimates. Hence, the average estimated errors of stellar population
parameters based on the observational errors are about $\pm0.1$ dex
for the log(age), [Z/H] and [E/Fe] and $\pm0.05$ for
log($\sigma$). 

The weak [O III] and/or H$\beta$ emission in the two galaxies
ESO318-G021 and MCG-02-13-009 (Hau \& Forbes 2006) is found to 
have no significant effect on our
measurements of the stellar population parameters, 
again because of our use of all available indices and the multi-index
$\chi^2$ minimization technique to fit the indices to the SSP model.

\section{results}
\subsection{Central stellar population parameters}
The central stellar population parameters of each galaxy are obtained
by averaging the 
values of all apertures within the central $r_e/8$. The values are
listed in Table 4 with the error on the mean.
In the following sections we will investigate these values and their
correlations compared to those for galaxies in high density
environments (HDEs). 

The stellar populations of three isolated galaxies NGC 821, NGC
1045 and NGC 2865 were previously measured in the literature.
Comparing our measurements of NGC 2865 to those of
S{\'a}nchez-Bl{\'a}zquez et al. (2007), we found comparable values of the
central age (1.7 vs. 1.0 Gyrs), [Z/H]$_o$ (0.48 vs. 0.52) and
[E/Fe]$_o$ (0.07 vs. 0.13).
On the other hand, comparing our measurements for NGC 1045 to those
obtained by using the data of Denicol{\'o} et al. (2005a), their data
indicate a younger age (5 vs. 10 Gyrs), more metal-rich (0.43
vs. 0.31) and less enhanced 
[E/Fe]$_o$ (0.12 vs. 0.36). We note that this galaxy does not show
emission lines and Denicol{\'o} et al. did not apply any emission
corrections to its indices. For NGC 821, both measurements of Proctor
et al. (2005) and those we obtained by using the data of Denicol{\'o} 
et al. (2005a) are comparable (see Table 2).
Through out this paper we will consider our measurements for NGC 1045
and NGC 2865 and Proctor et al. values for NGC 821. 

\subsubsection{Central values}

\begin{table*}
\begin{center}
\begin{tabular}{lcccc}
\multicolumn{5}{l}{\bf Table 4. \small Central (r$<$r$_e/8$) stellar
  population parameters and the error on the mean.} \\ 
\hline
 Galaxy         &  log($\sigma_o$)   &  age$_o$ (Gyr)  & [E/Fe]$_o$   &     [Z/H]$_o$  \\
 \hline  			       		  		
 NGC 682        &  2.30 $\pm$  0.01  &  8.0  $\pm$ 0.6 &  0.30 $\pm$ 0.01 &  0.33 $\pm$  0.02 \\
 NGC 1045       &  2.42 $\pm$  0.01  & 10.4  $\pm$ 0.6 &  0.36 $\pm$ 0.02 &  0.31 $\pm$  0.02 \\
 NGC 1162       &  2.29 $\pm$  0.01  &  6.8  $\pm$ 0.4 &  0.31 $\pm$ 0.01 &  0.29 $\pm$  0.02 \\
 NGC 2271       &  2.37 $\pm$  0.01  & 11.5  $\pm$ 0.5 &  0.14 $\pm$ 0.01 &  0.44 $\pm$  0.02 \\  
 NGC 2865       &  2.25 $\pm$  0.02  &  1.7  $\pm$ 0.1 &  0.07 $\pm$ 0.01 &  0.48 $\pm$  0.05 \\ 
 NGC 4240       &  2.09 $\pm$  0.01  &  7.4  $\pm$ 0.5 & --0.07 $\pm$ 0.01 &  0.23 $\pm$  0.02 \\ 
 ESO153-G003    &  2.34 $\pm$  0.01  & 11.2  $\pm$ 0.4 &  0.28 $\pm$ 0.01 &  0.19 $\pm$  0.01 \\
 ESO218-G002    &  2.43 $\pm$  0.01  & 14.8  $\pm$ 0.1 &  0.26 $\pm$ 0.01 &  0.35 $\pm$  0.01 \\
 ESO318-G021    &  2.37 $\pm$  0.01  &  8.9  $\pm$ 0.9 &  0.08 $\pm$ 0.01 &  0.34 $\pm$  0.03 \\
 MCG-01-27-013  &  2.39 $\pm$  0.01  &  8.0  $\pm$ 1.6 & --0.03 $\pm$ 0.01 &  0.34 $\pm$  0.04 \\
 MCG-02-13-009  &  2.33 $\pm$  0.02  &  9.6  $\pm$ 1.2 &  0.05 $\pm$ 0.02 &  0.36 $\pm$  0.02 \\
 MCG-03-26-030  &  2.47 $\pm$  0.01  & 14.5  $\pm$ 0.2 &  0.15 $\pm$ 0.03 &  0.27 $\pm$  0.01 \\
\hline 

\end{tabular}
\end{center}
\end{table*}

Including the additional data from Table 2, the seventeen
isolated galaxies show a range of central luminosity-weighted
ages. Roughly half (9 out of 17) are older than 9 Gyr, five galaxies
are between 5 and 8 Gyrs, and 3 with ages younger than 3 Gyrs.
The sample has an average age of $8.6\pm0.9$ Gyrs. This is
younger than the average age of about 11 Gyr for galaxies in the
Coma cluster found by Mehlert et al. (2003). 
Younger average luminosity-weighted ages for galaxies in low density
environments were also found by Proctor et al. (2004; see also
references therein) in comparison of galaxies in high density
environments (HDEs) such as clusters and Hickson compact 
groups to those in loose groups and field. Similar results were also
found in the large sample study of Terlevich \& Forbes (2002).

On average, our isolated galaxies have central luminosity-weighted
total metallicities of 
[Z/H]$_o =0.29\pm0.03$. Mehlert et al. (2003) quotes values of
$0.24\pm0.06$ and $0.12\pm0.17$ for the Coma
cluster galaxies of types E and E/S0 respectively and Collobert et
al. (2006) found an average of [Z/H]$_o =0.27$ for their cluster
galaxies. Using the  published stellar population parameters in Thomas
et al. (2005), we find that the early-type galaxies in their HDE
subsample have an average of [Z/H]$_o=0.29\pm0.02$, while their
galaxies in low density environments are more metal-rich by about
0.05-0.1 dex. 
 
The average central luminosity-weighted [E/Fe]$_o$ of our isolated
galaxies is $0.17\pm0.03$.
Comparing to the mean value of [E/Fe]$_o\sim0.26\pm0.06$ 
reported by Mehlert et al. (2003) for
cluster galaxies, our isolated galaxies show lower values by order of
0.1 dex.
Although, if we arbitrarily divide our sample in half, then 8 out of
17 of our galaxies are significantly
enhanced with $0.20<$[E/Fe]$<0.36$. These eight galaxies are comparable
to those measured by Proctor et al. (2004) for galaxies in massive and
compact groups ([E/Fe]$_o=0.26\pm0.04$). 
The other nine galaxies resemble the solar element abundance
with [E/Fe]$_o\sim 0.06\pm0.03$ which is lower than what Proctor et
al. obtained for field galaxies ([E/Fe]$=0.12\pm0.02$).

\subsubsection{Central parameters correlations}

\begin{figure*}
\centerline{\psfig{figure=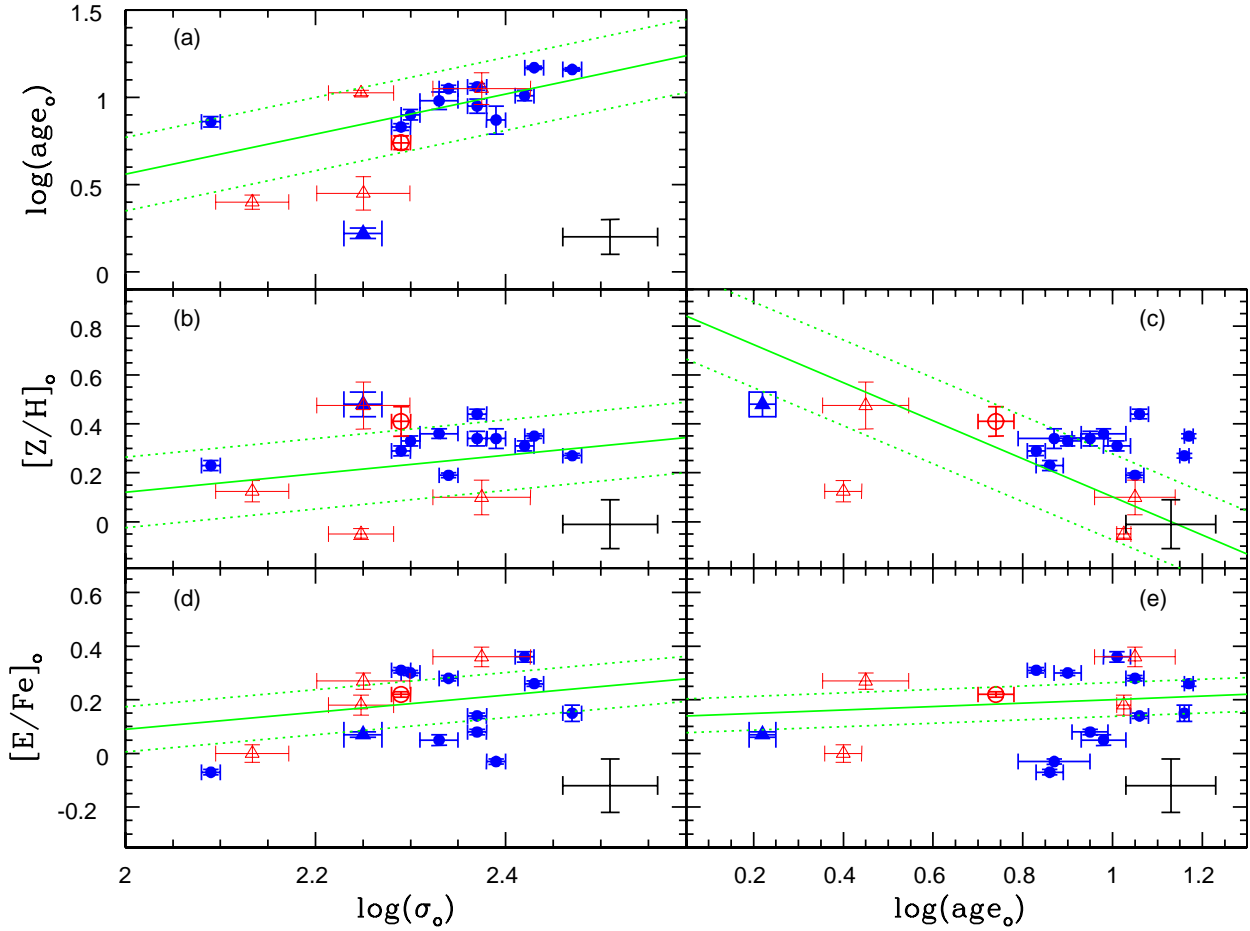,width=0.75\textwidth,angle=-90}}
\caption
{
 Relations between the central age, velocity dispersion ($\sigma_o$), total
metallicity [Z/H]$_o$ and $\alpha$-elements abundance
[E/Fe]$_o$. 
Solid lines are correlations from Bernardi et al. (2006) for 
low redshift $\le0.06$ galaxies in HDEs, and
dotted lines represent their $1\sigma$ scatter.
In panel (a), isolated galaxies exhibit slightly more scatter
towards younger ages for their velocity dispersion than
those in dense environments.
In panel (b), the [Z/H]$_o$ and velocity dispersion of isolated
galaxies follow a similar trend as the HDE galaxies with a tendency to
 higher metallicities.  
In panel (c), isolated galaxies follow a similar age-metallicity
 relation as those in HDE with a tendency to higher metallicities.
In panels (d) and (e), isolated galaxies exhibit more scatter
towards lower [E/Fe]$_o$  for their velocity dispersion and age than
those in HDEs.
In all panels, solid symbols are data of the present sample, the solid
triangle  represents NGC 2865, the open circle is NGC 821 from Proctor
et al. (2005) and open triangles are data for galaxies from Denicol{\'o}
et al. (2005a). We plot data for NGC 1045 from our present study only.
Error bars in the bottom right corners are the  
systematic error of $\pm0.1$ dex for age, [Z/H]$_o$, [E/Fe]$_o$ and
$\pm0.05$ for log($\sigma_o$).  
}
\label{centr}
\end{figure*}
Above we have compared the central stellar population parameters of the
isolated galaxies with their counterparts in HDEs. However, such
parameters are known to vary with galaxy velocity dispersion
(e.g. Kuntschner et al. 2001, 2002; Mehlert et al. 2003;  Collobert et
al. 2006; S{\'a}nchez-Bl{\'a}zquez et al. 2006b; Brough et al. 2006). 
Next we compare the correlations between central stellar population
parameters of our isolated galaxies to those in HDEs. 

Bernardi et al. (2006) identified a sample of 490 early-type
galaxies in HDEs from the SDSS at redshifts of $z\le0.06$. They define the 
HDEs as those galaxy systems of total luminosity more than 10 times
the luminosity of a typical early-type galaxy.
To obtain a high S/N ratio for their spectra, they co-added similar
objects by using narrow bins in luminosity, size, velocity
dispersion, and redshift (Bernardi et al. 2003). The 490 galaxies
produced 105 composite spectra of S/N$\sim 100$.
Using their published stellar population parameters and 
correlations between those 
parameters, we have measured the $1\sigma$ scatter of their galaxies about
these correlations. In Fig. \ref{centr}, solid lines are the best fit
relation from Bernardi et al., and
dotted lines represent the $1\sigma$ scatter calculated by us.
Similar trends to Bernardi et al. were found for galaxies in HDEs by
previous studies (e.g. Kuntschner et al. 2002; Mehlert et al. 2003;
 Collobert et al. 2006; S{\'a}nchez-Bl{\'a}zquez et al. 2006b). 

For our isolated galaxies, trends between the stellar population
parameters in the central regions and the central velocity
dispersion (a proxy for mass) are 
examined in Fig. \ref{centr} (a),(b) and (d), while panels (c) and (e)
show correlations of [Z/H] and [E/Fe] with age. Isolated galaxies
follow similar trends to those in HDEs with more massive galaxies
being older, more metal-rich and with higher [E/Fe]$_o$ than less
massive ones. Isolated galaxies show slightly higher scatter towards
younger ages, higher [Z/H]$_o$ and 
lower [E/Fe]$_o$ for their velocity dispersion than those in HDEs.
In panels (d) and (e), the two galaxies NGC 4240 and MCG-01-27-013 are
the most deviant galaxy from the correlations with very low [E/Fe]$_o$
for their velocity dispersion and age.

Similarly, comparing galaxies in low density environments with those
in Fornax cluster, Kuntschner et al. (2002) found that both samples
follow similar log($\sigma_o$)-metallicity trends, although galaxies
in low density environments show higher metallicities by $\simeq0.15$
dex.

Panel (c) shows the observed age-metallicity relation for the isolated
galaxies. 
Although the correlated errors of the age and metallicity may be
partly responsible for this correlation (e.g. Kuntschner et al. 2001),
our isolated galaxies follow a similar correlation to HDE galaxies.
While the effect of the correlated errors is expected to reduce the
measured metallicity of
old galaxies, we note that the oldest three of our galaxies (ESO218-G002,
MCG-03-26-030 and NGC 2271) with age $>11$ Gyrs, tend
to be more metal-rich than the 
average galaxy. On the other hand, the galaxy NGC 6172 tends to be younger
and less metal-rich than the general trend of the correlation. The
stellar population of this galaxy is extracted from the emission
corrected Lick absorption indices from Denicol{\'o} et al. (2005a), which
perhaps leads to measuring a younger age.

As a function of central age, panel (e) shows that many of the
isolated galaxies have low 
[E/Fe]$_o$ for their age compared to their counterparts in HDEs.
Low values of [E/Fe]$_o$  have been previously reported by
Collobert et al. (2006) for their galaxies in low density environments.

In summary, isolated galaxies in our sample span a large range of
ages. 
Scaling relations between central stellar parameters are similar to
those for galaxies in higher density environments albeit with a
tendency to younger central ages, higher [Z/H] and lower [E/Fe].

\begin{figure*}
\centerline{\psfig{figure=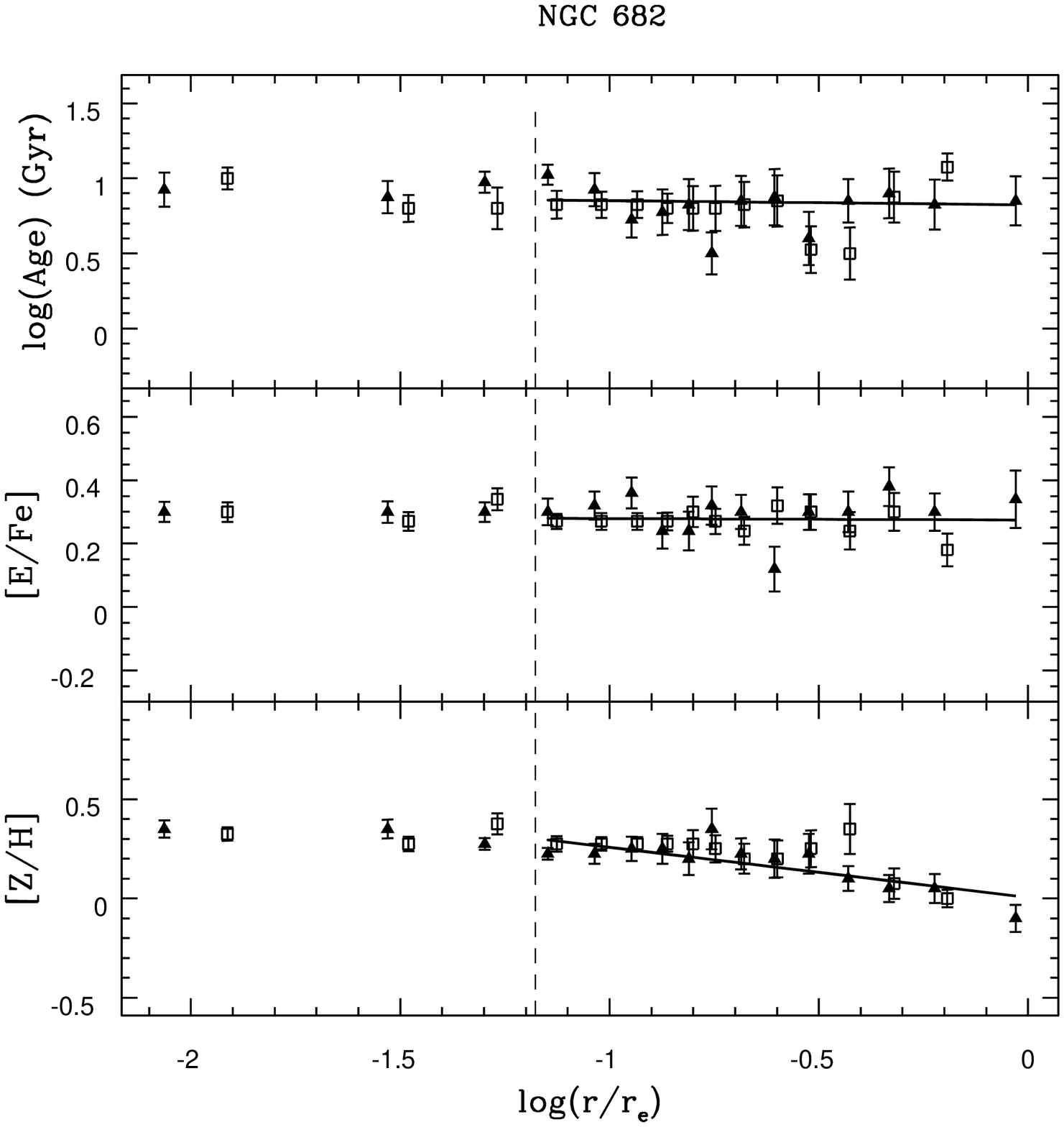,width=0.47\textwidth}
\psfig{figure=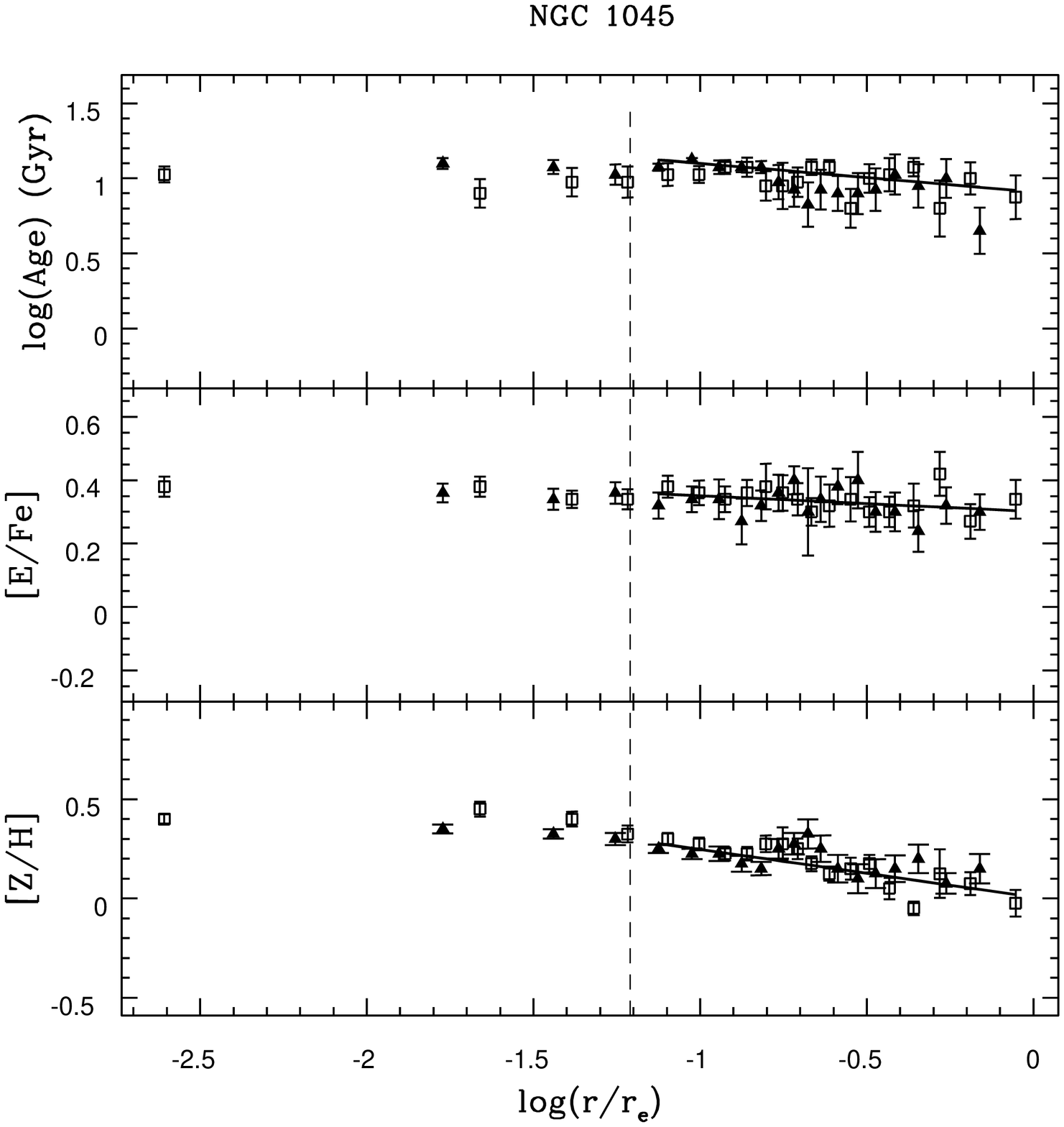,width=0.47\textwidth}}
\centerline{\psfig{figure=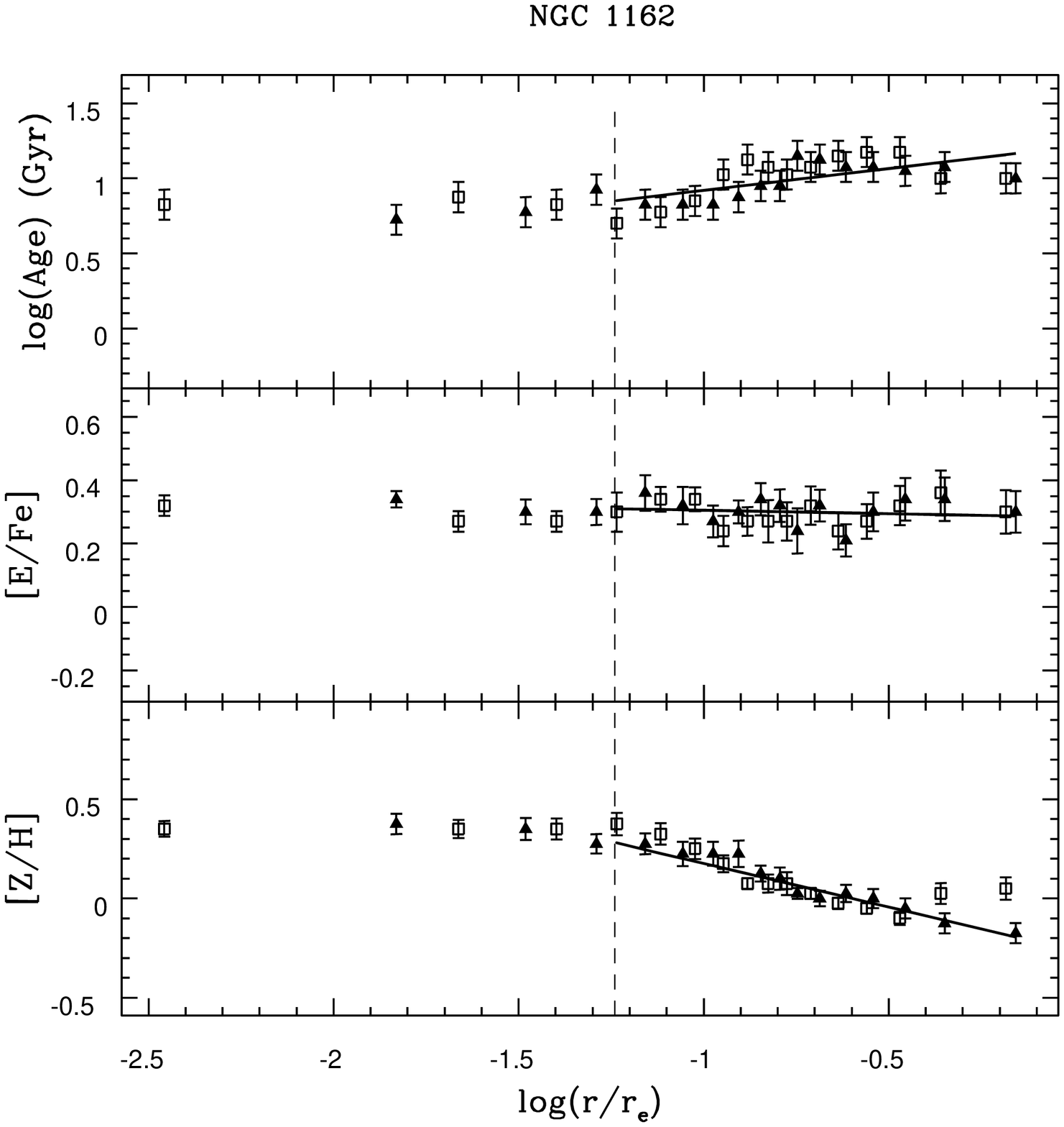,width=0.47\textwidth}
\psfig{figure=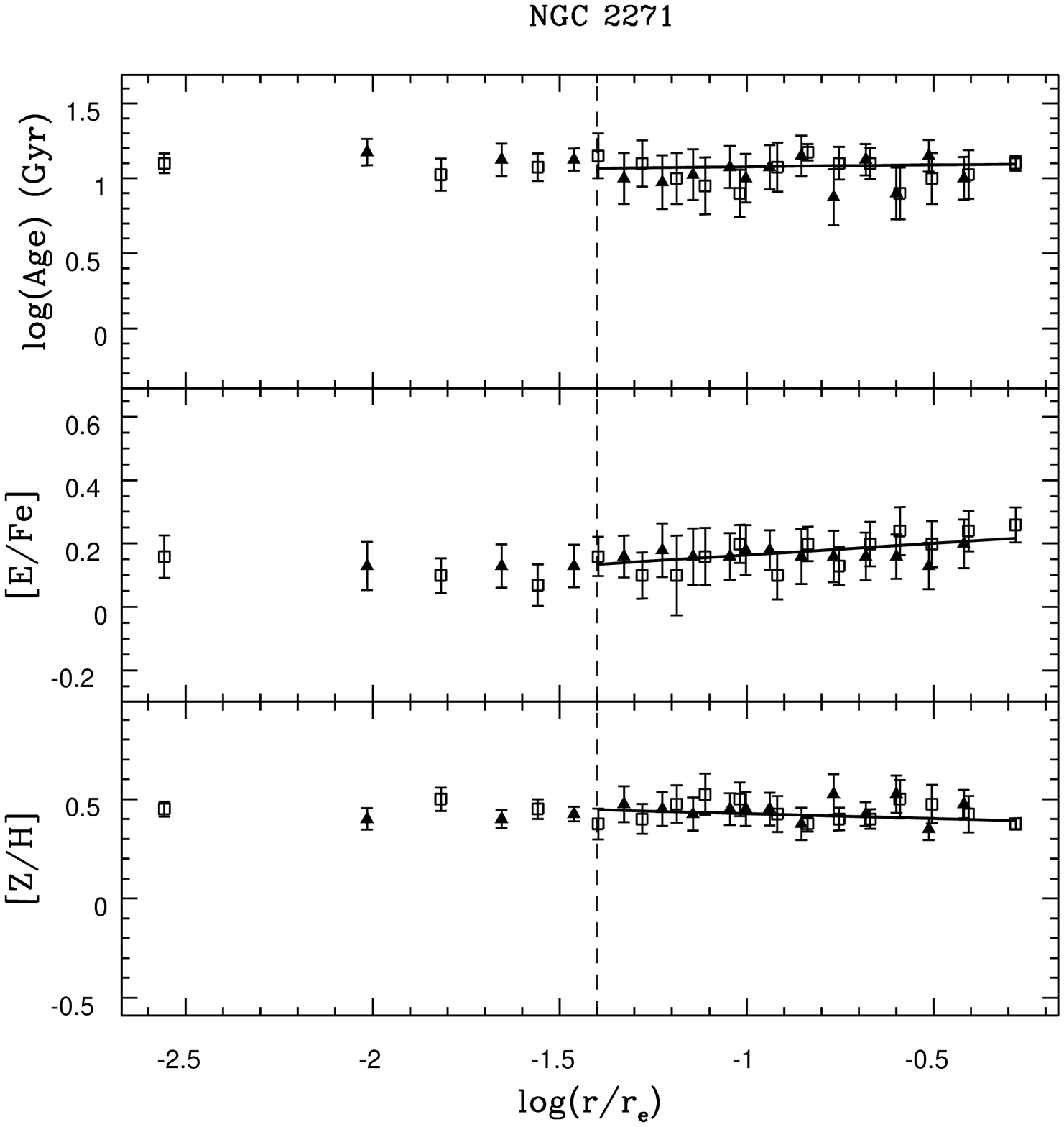,width=0.47\textwidth}}
\caption
{
The logarithmic radial profile of log (age), [E/Fe] and  [Z/H]
in terms of the effective radius. Squares and
triangle symbols represent the apertures on different sides of the
galaxy centre. The vertical dashed line represents the seeing limit.
The solid line is the weighted linear least square fit to all
apertures beyond the seeing limit. 
}
\label{test1}
\end{figure*}

\begin{figure*}
\centerline{\psfig{figure=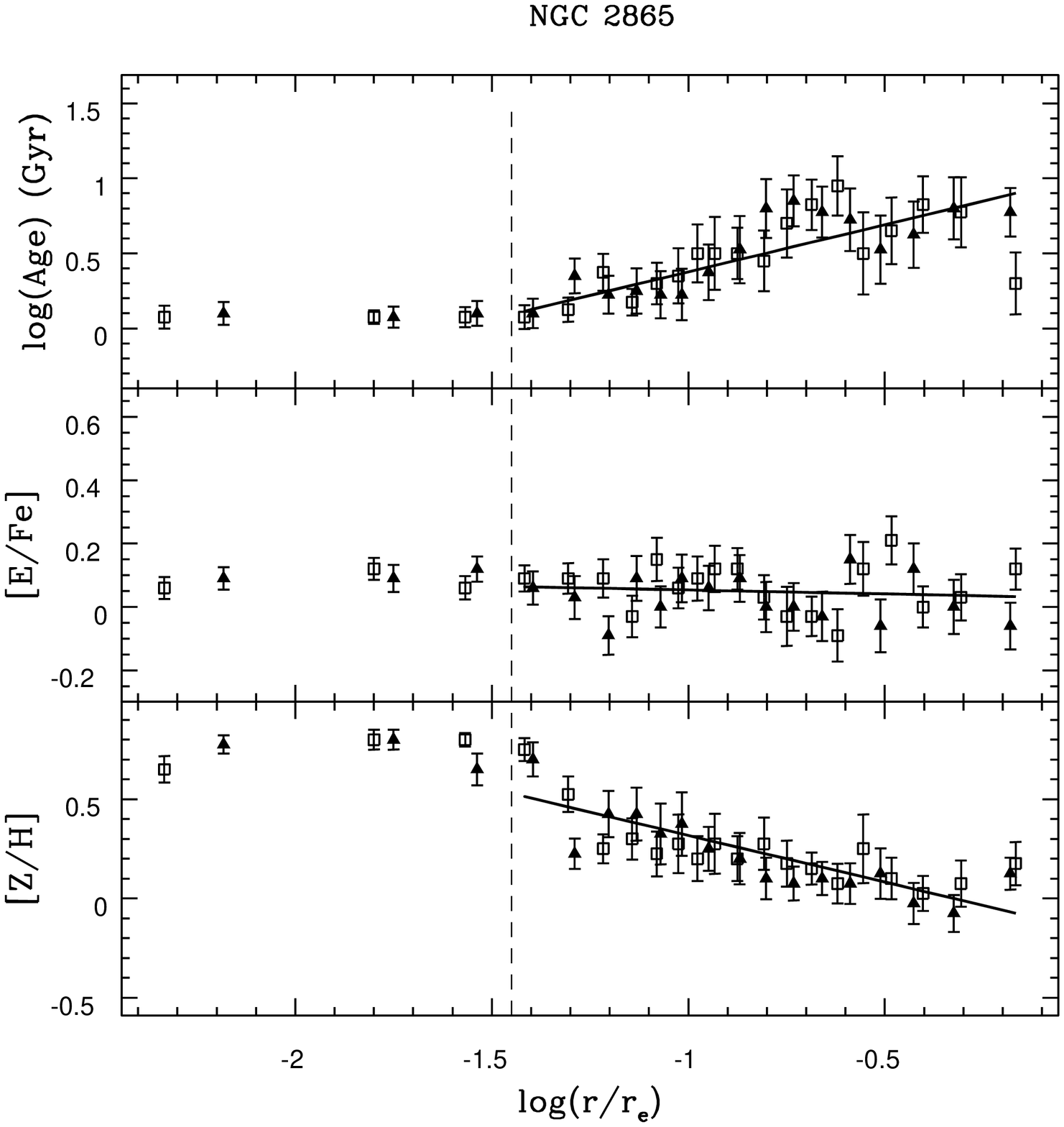,width=0.47\textwidth}
\psfig{figure=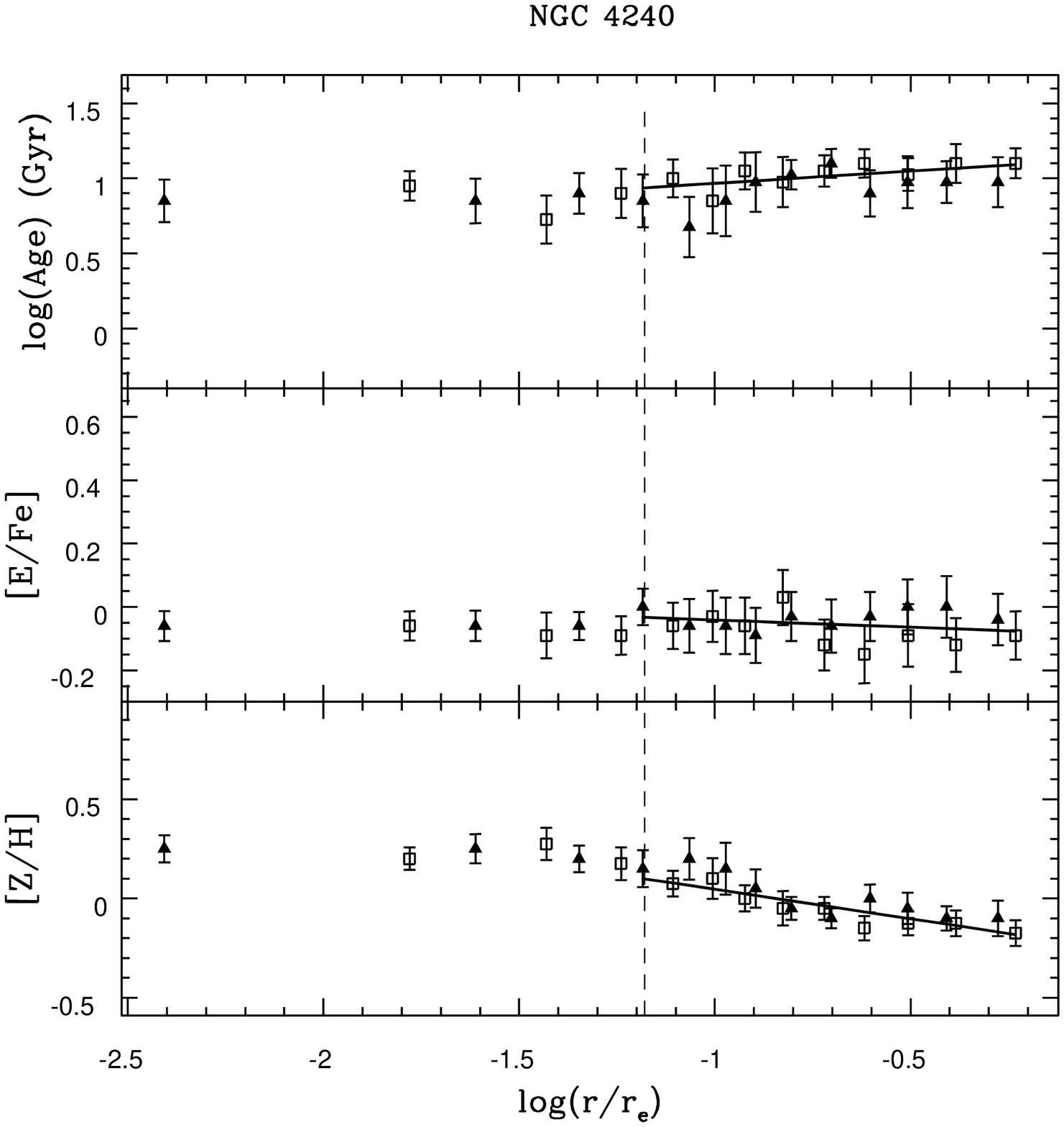,width=0.47\textwidth}}
\centerline{\psfig{figure=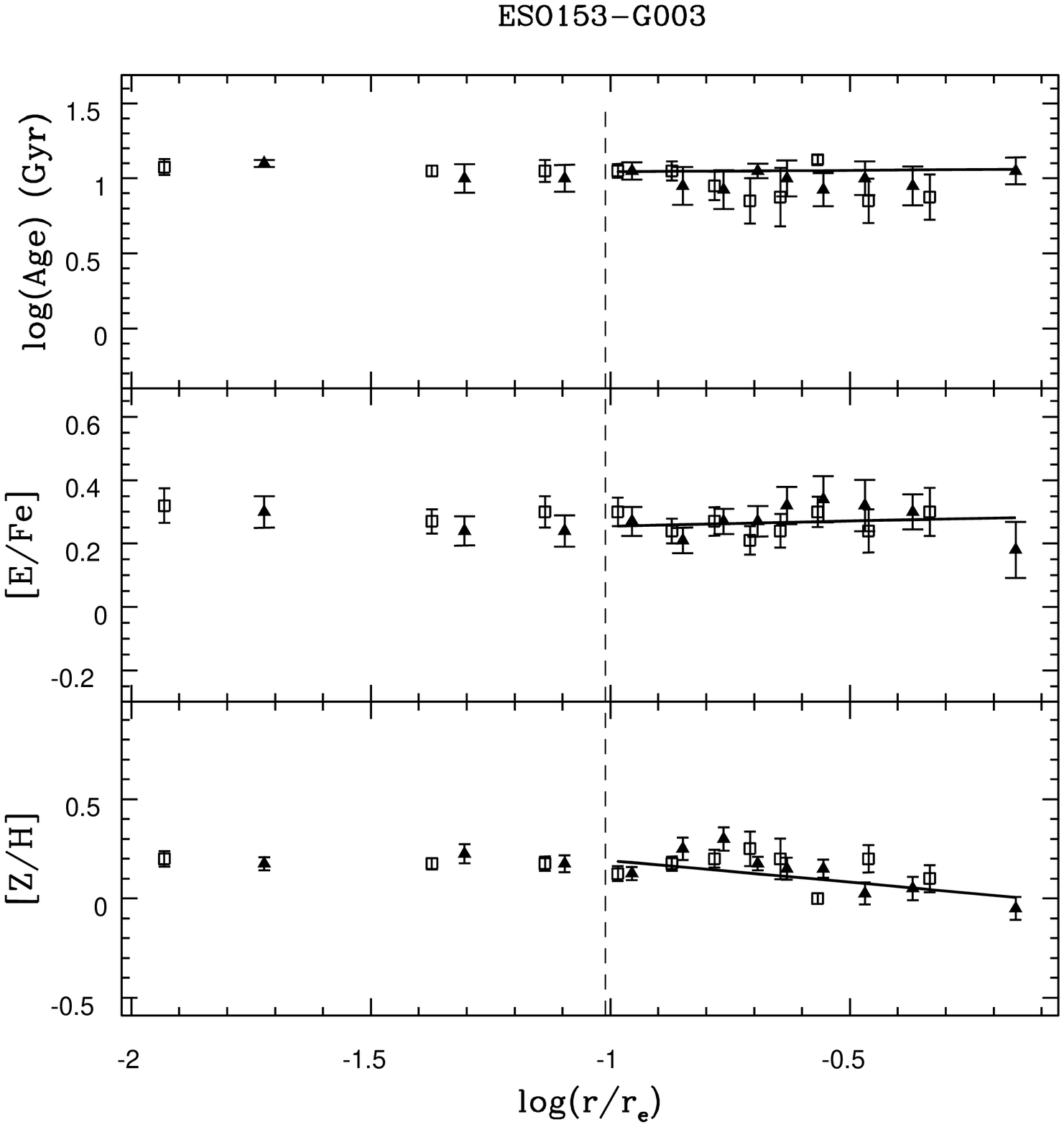,width=0.47\textwidth}
\psfig{figure=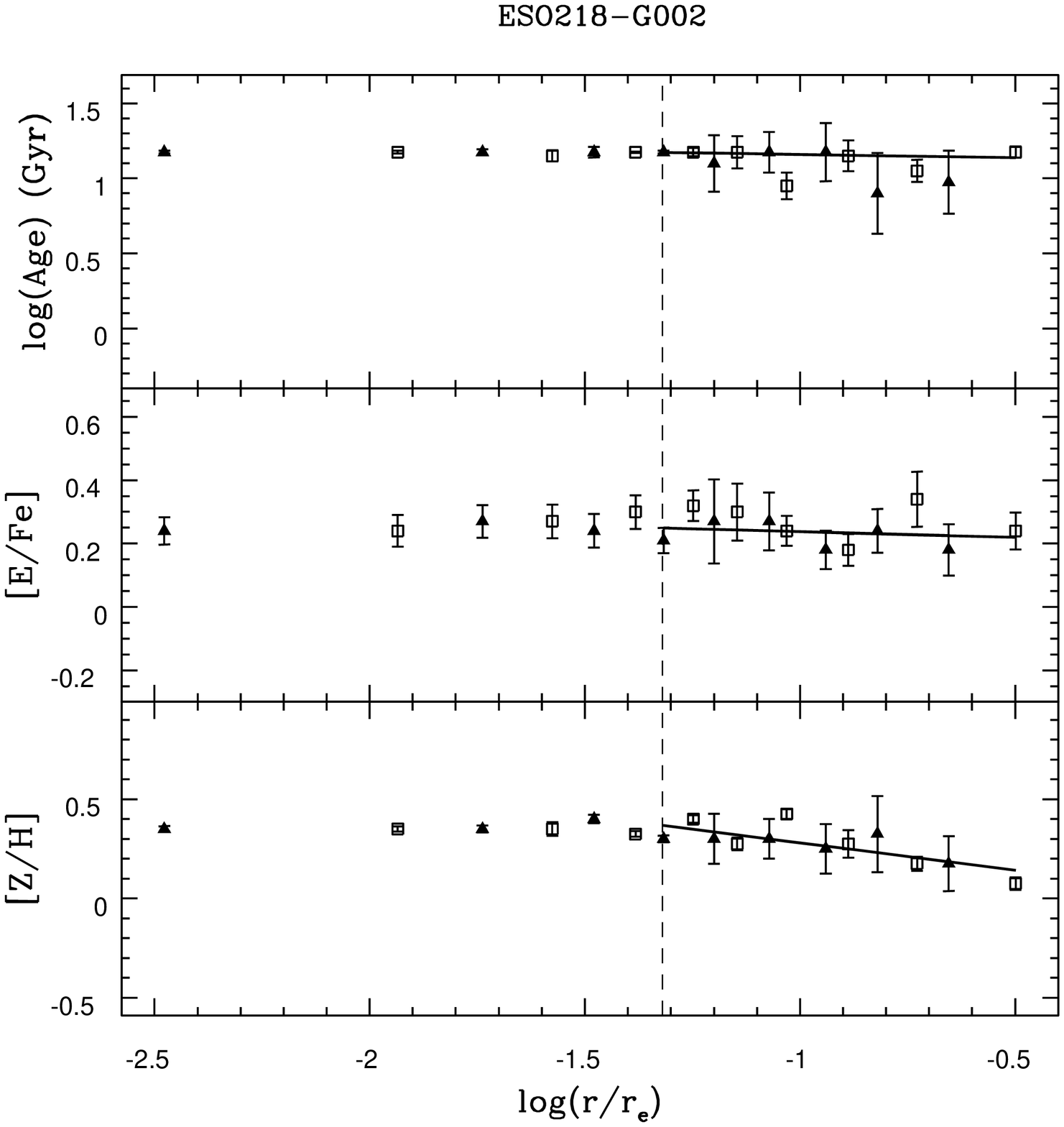,width=0.47\textwidth}}
\contcaption{
}
\label{test2}
\end{figure*}

\begin{figure*}
\centerline{\psfig{figure=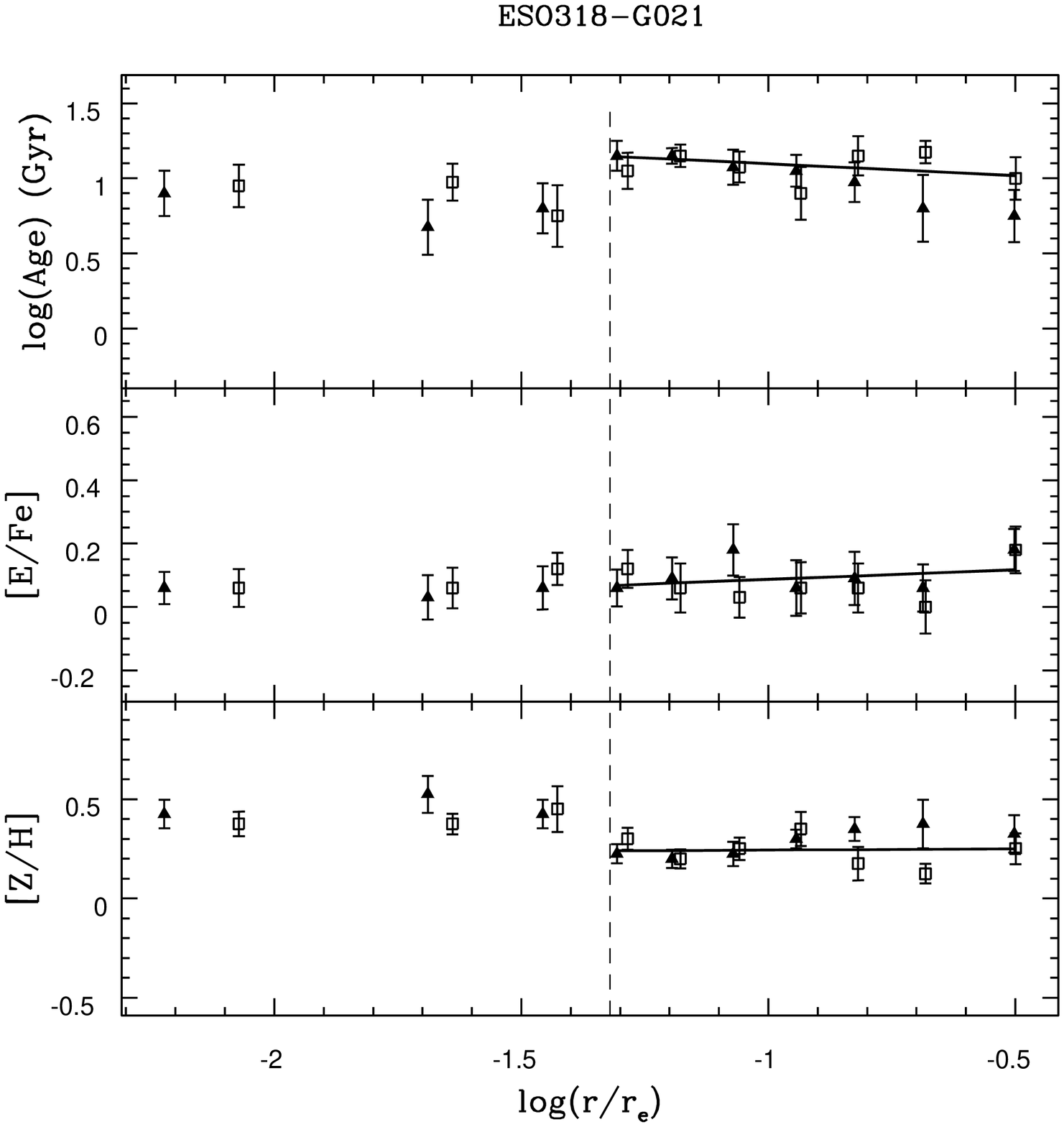,width=0.47\textwidth}
\psfig{figure=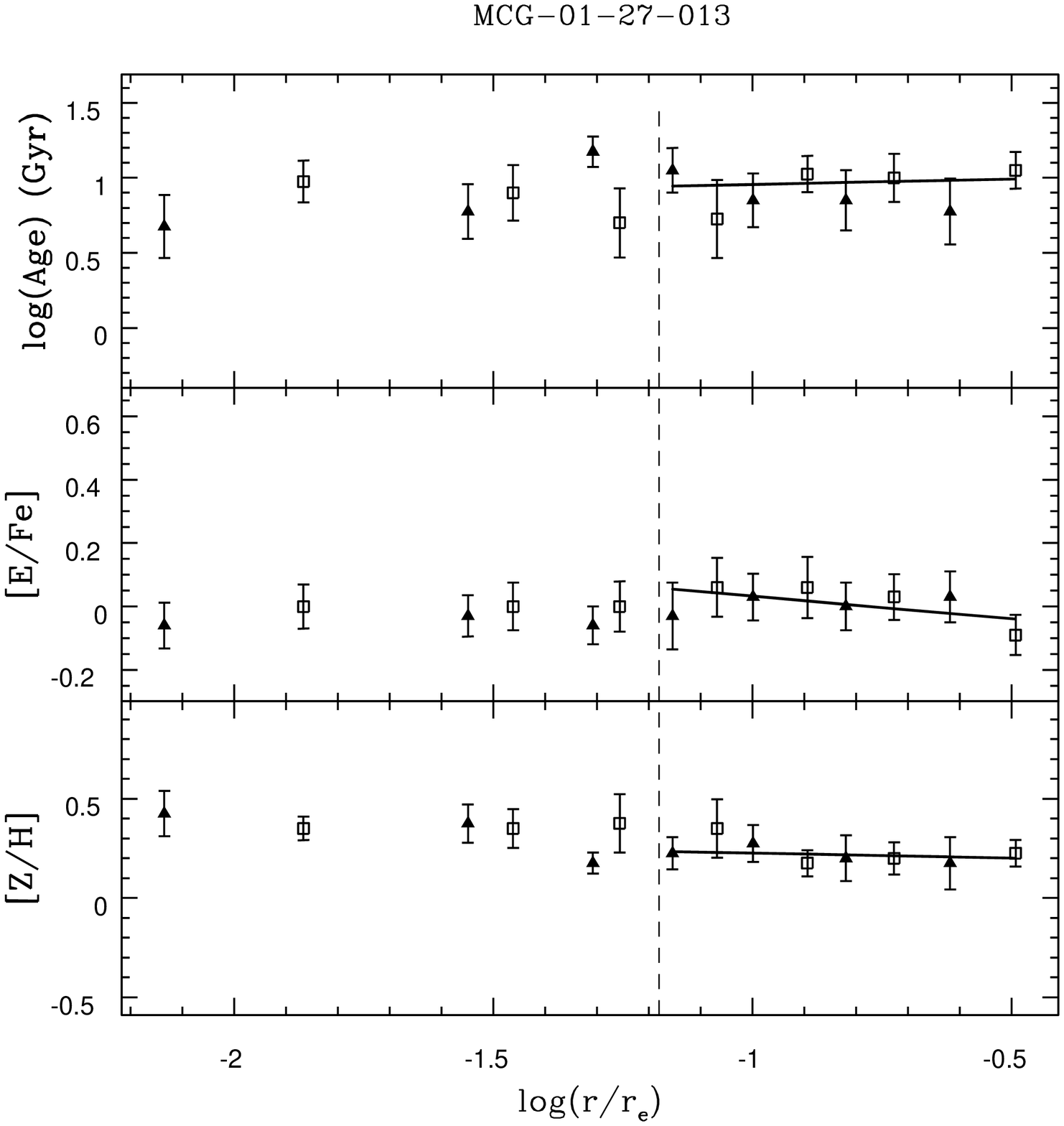,width=0.47\textwidth}}
\centerline{\psfig{figure=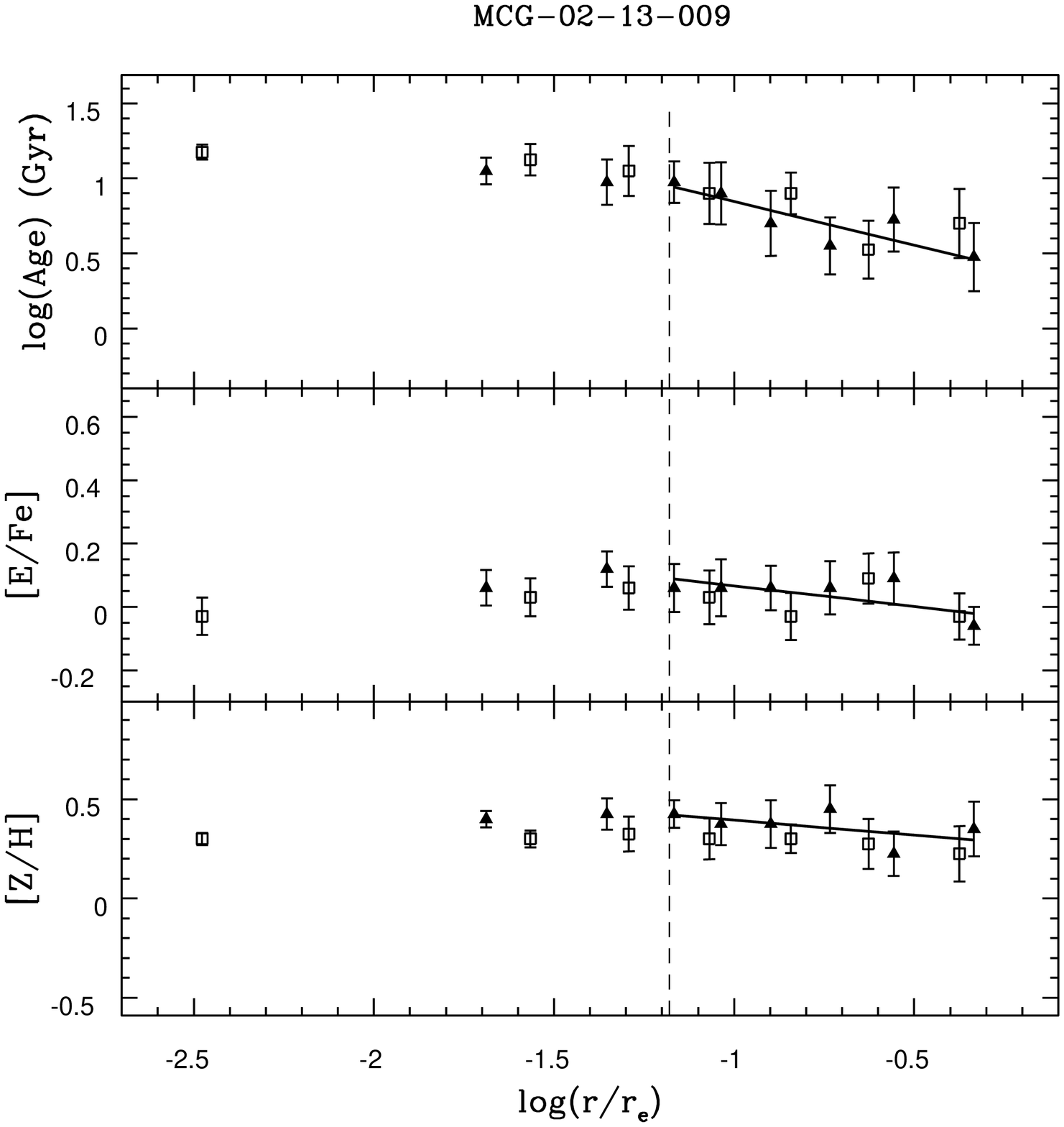,width=0.47\textwidth}
\psfig{figure=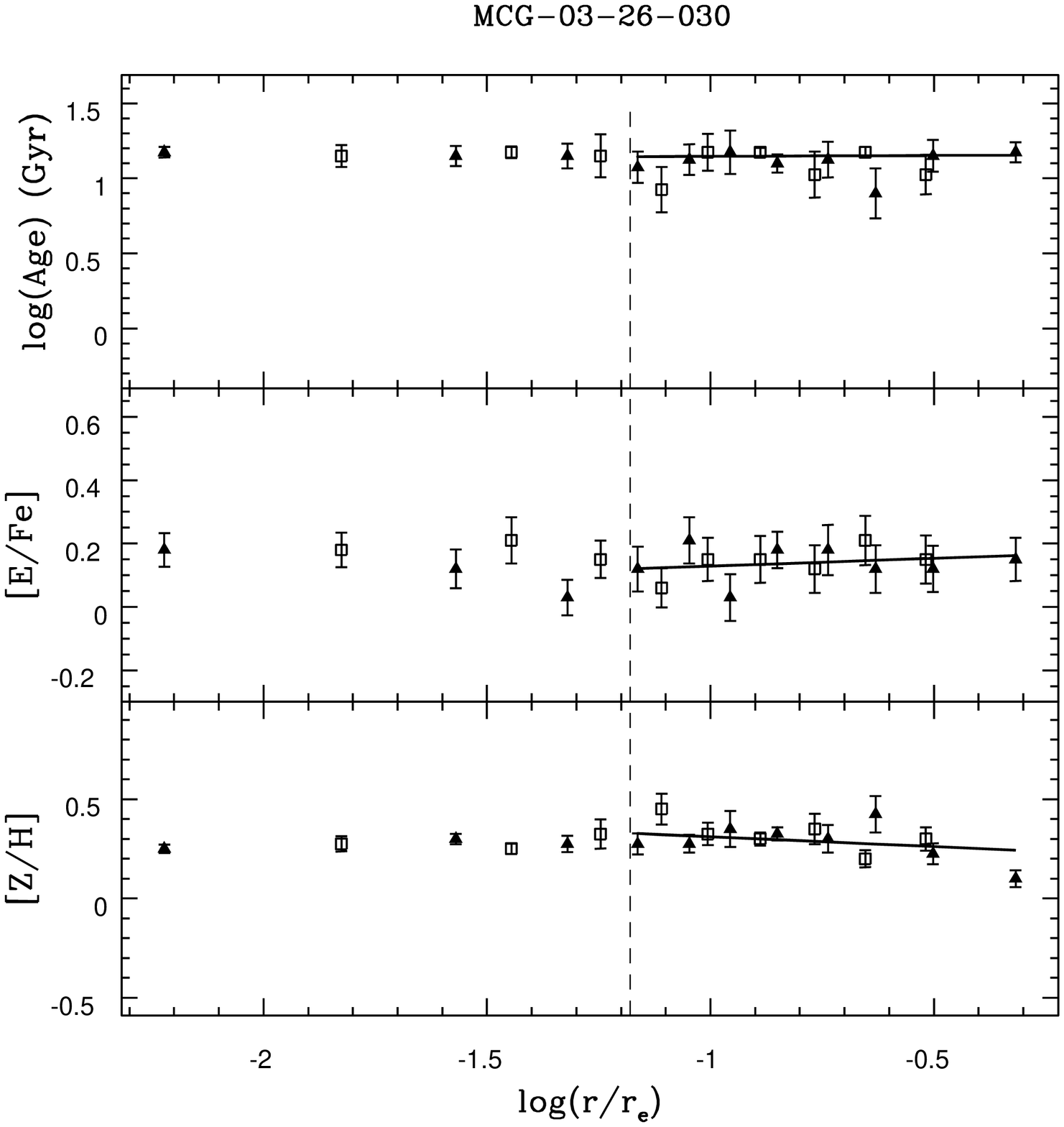,width=0.47\textwidth}}
\contcaption
{
}
\label{test3}
\end{figure*}

\subsection{Kinematic and stellar population radial profiles}
The radial recession profile of the stellar population parameters are fit
with a weighted least squares method to measure the
gradient for  all apertures beyond the seeing limit of $1''$. These
fits are listed in Table 5.  
We note that our measurements of 
[Z/H] and [E/Fe] gradients of the galaxy NGC 2865 are comparable to those
found by S{\'a}nchez-Bl{\'a}zquez et al. (2007), although they found a steeper
age gradient of $1.15\pm0.05$. Throughout this paper we use our
measurements for this galaxy which is shown in all plots as a solid
triangle. 

In the following sections we examine the radial profile of each 
 parameter in detail. 

\begin{table*}
\begin{center}
\begin{tabular}{lccc}
\multicolumn{4}{l}{\bf Table 5. \small The radial gradients of the
  age, [E/Fe]  and [Z/H] parameters.} \\
\hline
 Galaxy       &    log(age)  &     [E/Fe]   &     [Z/H]   \\     
 \hline    
NGC 682       & --0.03 $\pm$ 0.08 & --0.01 $\pm$ 0.03 & --0.25 $\pm$ 0.03  \\
NGC 1045      & --0.19 $\pm$ 0.04 & --0.05 $\pm$ 0.03 & --0.24 $\pm$ 0.03  \\
NGC 1162      &   0.29 $\pm$ 0.07 & --0.02 $\pm$ 0.04 & --0.44 $\pm$ 0.03  \\
NGC 2271      &   0.03 $\pm$ 0.07 &   0.07 $\pm$ 0.05 & --0.05 $\pm$ 0.04  \\
NGC 2865      &   0.63 $\pm$ 0.07 & --0.03 $\pm$ 0.03 & --0.47 $\pm$ 0.05  \\
NGC 4240      &   0.16 $\pm$ 0.12 & --0.01 $\pm$ 0.06 & --0.30 $\pm$ 0.06  \\
ESO153-G003   &   0.02 $\pm$ 0.08 &   0.03 $\pm$ 0.07 & --0.22 $\pm$ 0.05  \\
ESO218-G002   & --0.04 $\pm$ 0.04 & --0.04 $\pm$ 0.07 & --0.28 $\pm$ 0.04  \\
ESO318-G021   & --0.16 $\pm$ 0.11 &   0.06 $\pm$ 0.07 &   0.01 $\pm$ 0.07  \\
MCG-01-27-013 &   0.07 $\pm$ 0.25 & --0.14 $\pm$ 0.13 & --0.05 $\pm$ 0.14  \\
MCG-02-13-009 & --0.58 $\pm$ 0.27 & --0.13 $\pm$ 0.10 & --0.15 $\pm$ 0.15  \\
MCG-03-26-030 &   0.01 $\pm$ 0.04 &   0.05 $\pm$ 0.08 & --0.10 $\pm$ 0.02  \\
\hline 
\end{tabular}
\end{center}
\end{table*}

\subsubsection{Rotation velocity and velocity dispersion profiles} 

The detailed kinematics of our sample of galaxies including the radial
velocity and velocity dispersion profiles of the galaxies have been
previously studied by Hau \& Forbes (2006). Reproducing the velocity
profiles (see Sec 2.3), we confirm their findings of kinematic substructures in some
of these galaxies. Rapidly rotating cores are detected in seven
galaxies, three reveal fast rotating outer discs, while three galaxies
show central peaked velocity dispersion profiles (see Hau \& Forbes
for details).

\subsubsection{Age profiles}

Although having a wide range of ages, Figs \ref{test1} show that the
majority of our sample of isolated early-type galaxies show
statistically insignificant age gradients with an average of
$0.04\pm0.08$ dex per dex. 
Four of these galaxies (NGC 2271, ESO153-G003, ESO218-G002 and
MCG03-26-030) have uniform ages older than 10 Gyrs indicating a
formation epoch at $z>2$. Two galaxies, ESO318-G021 and
MCG01-27-013, show intermediate ages of 8.9 and 8.0 Gyrs which
indicate an extended star formation history to $z\approx 1$.  

Although two galaxies, NGC 682 and NGC 4240, show insignificant age
gradients, they are composed of both old and relatively younger
populations, which suggests a secondary 
star formation epoch for these two galaxies. 
NGC 682 shows a global age of $\sim 8.0\pm0.6$ Gyrs, with a younger
population of about $4.0\pm0.5$ Gyrs in the region around 
$\log$($r/r_e$)$ = -0.5$.
The stellar kinematics at the same region and outwards
indicates the presence of a rotating disk with a rotational velocity of
150 km/s combined with a low velocity dispersion which declines
from about 200 km/s at $r/r_e = 0.15$ reaching 130 km/s at
the outer regions. For NGC 4240, despite its statistically
insignificant age gradient
of $+0.16 \pm 0.12$, it has evidence for  younger stars
($\sim 7.4\pm0.5$ Gyr) in the central regions at $r/r_e<0.1$. 

Remarkable positive gradients, with the central
regions being younger, are seen in two galaxies. 
The galaxy NGC 1162  shows a gradient of $+0.29\pm 0.07 $ with the
central stellar population having an
intermediate age of about $6.8\pm0.4$ Gyr for $r/r_e < 0.13$ compared
to an age $>10$ Gyr in the outer regions.  NGC 2865 is
the youngest member of our sample 
with a central stellar population as young as 1.7 Gyr for $ r/r_e <
0.2$ and intermediate age stars of $\sim7$ Gyr outwards. This
galaxy shows the steepest overall age gradient of $+0.63 \pm 0.07$.
The kinematics of the central (r$< 4 ''$) regions of these two
galaxies suggest rotationally supported cores (Hau \& Forbes 2006). 

NGC 1045 and MCG-02-13-009 show negative age gradients of $-0.19\pm0.04$
and $-0.58\pm0.27$ respectively. The former galaxy has a population of
intermediate age stars ($\sim 7$ Gyr) beyond $r/r_e \sim 0.2$. This
galaxy also shows  
asymmetrical rotation and a declining velocity dispersion profile outwards
indicating velocity substructure (see also Hau \& Forbes 2006). 
While the later galaxy shows a steady age decrease
outwards, a peaked central velocity dispersion and a slowly rotating body.

We note that in the HDE of the Coma cluster, Mehlert
et al. (2003) measured insignificant age gradients for 91 per cent of
their galaxies. 

\begin{figure}
\centerline{\psfig{figure=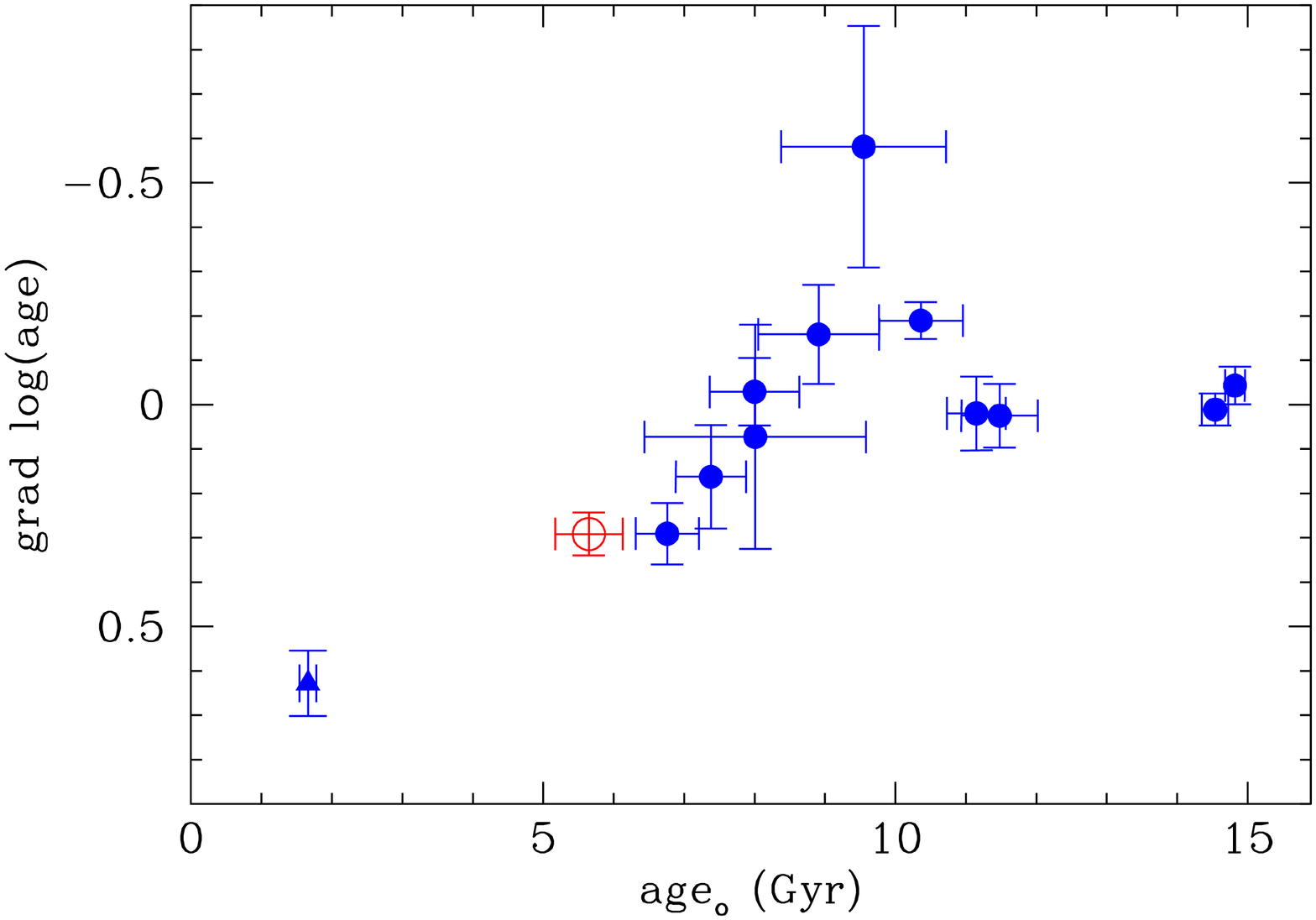,width=0.37\textwidth,angle=-90}}
\caption
{
The radial gradient of the age shows a strong 
correlation  with central values for galaxies younger than
$\sim 11$ Gyr,
while older galaxies show flat age profiles. Symbols as in Fig. \ref{centr}.
}
\label{grdage1}
\end{figure}

\subsubsection{Metallicity profiles}\label{z_profile}

The mean metallicity gradient of the 17 isolated galaxies is
$-0.25\pm0.05$.  
The two youngest galaxies of the sample NGC 1162 and NGC
2865, both have steep positive age gradients, and show steep
metallicity gradients of $-0.44 \pm 0.03$ and $-0.47\pm 0.05$
respectively. NGC 821 also has a young central stellar population and
a very steep [Z/H] gradient of $-0.72\pm0.04$ (Proctor et al. 2005).

On the other hand, three galaxies (ESO318-G021, MCG
-01-27-013 and MCG-02-13-009) which show statistically insignificant
 age gradients, also show negligible change of the 
metallicity between the centre and the outer regions with average radial 
gradients of $-0.06\pm0.03$. The flat metallicity gradient of these
three galaxies, can perhaps be due to the short coverage of our
spectra, i.e. only to r$_e/3$. 
While Mehlert et al. (2003) measure on average zero age gradients for
galaxies in the Coma cluster, the total
metallicity gradient was more pronounced (although with a large
scatter) with their galaxies having a mean negative gradient of
$-0.16\pm0.12$. S{\'a}nchez-Bl{\'a}zquez et al. (2007)
measure an average [Z/H] gradient of $-0.31\pm0.13$ for a sample of
galaxies in mostly HDEs.

\subsubsection{$\alpha$-element profiles}

Figs \ref{test1} show that our sample galaxies have insignificant 
radial gradients of [E/Fe] with an average of $-0.03\pm0.02$. 
Despite the measured flat gradient of [E/Fe] for the galaxy NGC 2865,
visual inspection reveals fluctuations between [E/Fe]=--0.1 and
0.1 which may be related to the shell structure of this galaxy.
Insignificant radial gradients of [E/Fe] were also found by Mehlert et
al. (2003) for the early-type galaxies in the Coma
cluster with an average of $0.05\pm0.05$.

\subsubsection{Radial gradient  correlations}

Although most of our galaxies reveal statistically insignificant age
gradients, Fig. \ref{grdage1} shows that while galaxies older than
11 Gyrs 
show no age gradients (i.e. uniformally old stellar populations), the age
gradients of young galaxies are strongly  anti-correlated with the
central age. 
No significant dependence of the age gradient on the central velocity
dispersion, [Z/H]$_o$ or [E/Fe]$_o$ was found.
This suggests that galaxy mass and star formation timescale plays
little role in establishing age gradients. Thus as a young starburst
evolves, the age gradient flattens from positive to almost zero.

\begin{figure*}
\centerline{\psfig{figure=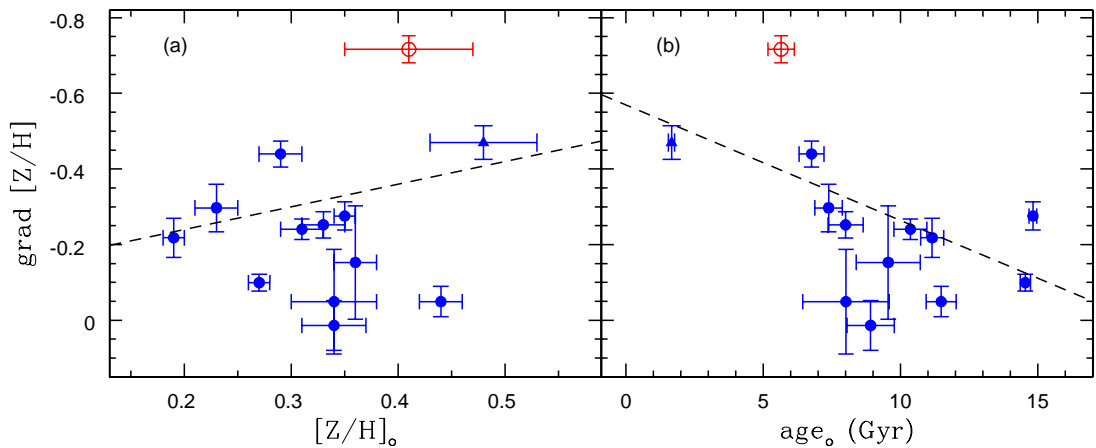,width=0.37\textwidth,angle=-90}}
\caption
{
The radial gradient of the total metallicity [Z/H] versus a) the central
total metallicity [Z/H]$_o$ and b) the central age. The dashed line
represent the trend found by 
S{\'a}nchez-Bl{\'a}zquez et al. (2007) for HDE galaxies. Excluding galaxies
with flat metallicity 
gradients (see text), our isolated galaxies follow similar
trends. Symbols as in Fig. \ref{centr}.
}
\label{grdz1}
\end{figure*}

\begin{figure}
\centerline{\psfig{figure=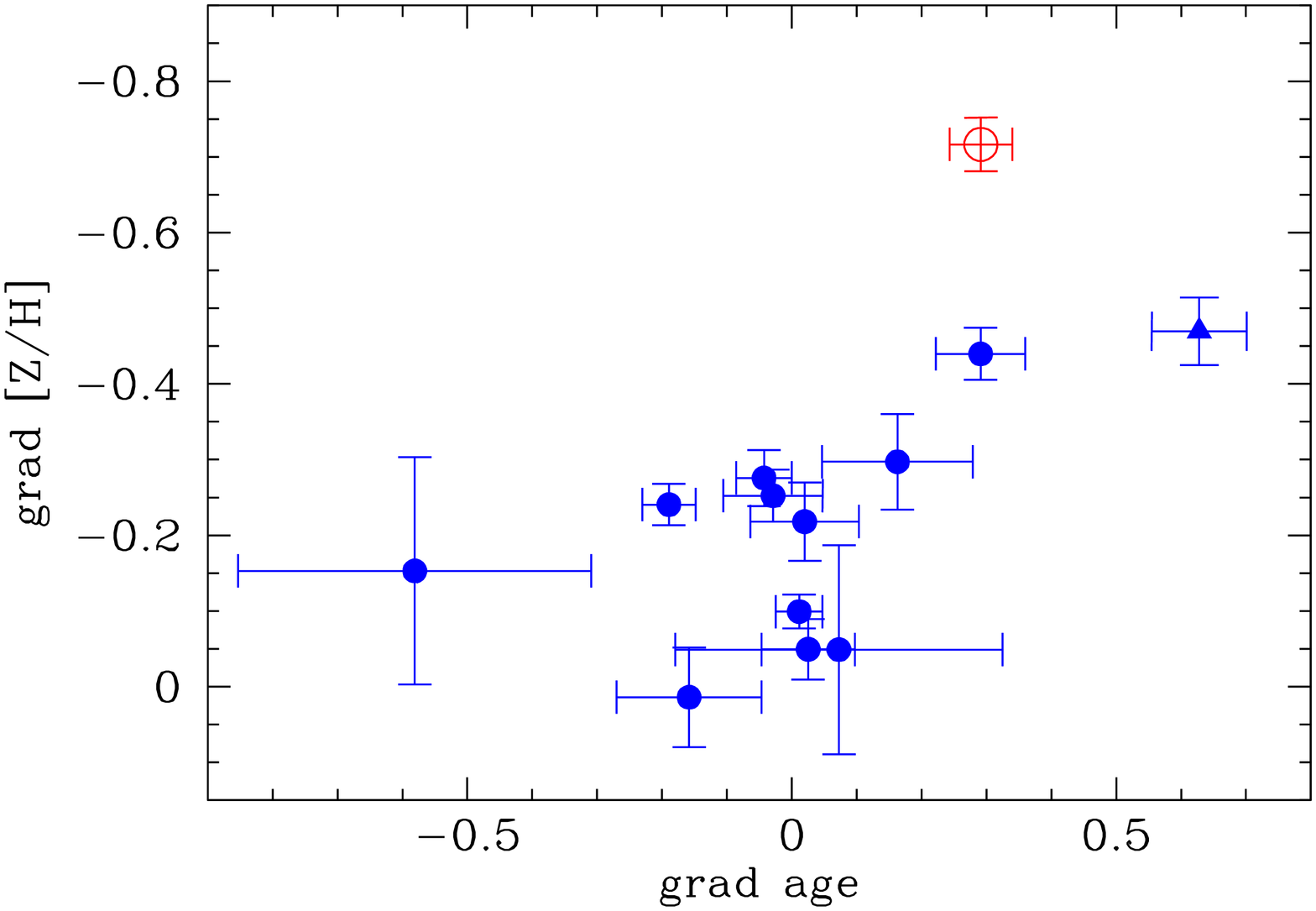,width=0.37\textwidth,angle=-90}}
\caption
{
The radial gradient of [Z/H] is strongly correlated to the age
gradient. Symbols as in Fig. \ref{centr}.  
}
\label{gzgage}
\end{figure}

Fig. \ref{grdz1} panel (a) shows the total metallicity gradient as
a function of its central value. 
The dashed line represents the trend found by
S{\'a}nchez-Bl{\'a}zquez et al. (2007) for galaxies mostly in HDEs of 
clusters and groups. Excluding galaxies with flat metallicity
gradients (see Sec. \ref{z_profile}), our galaxies follow the trend.
This trend indicates that galaxies with steeper gradients have more
metal-enriched inner regions. 
The galaxy NGC 2271 is the most deviant from the relation with a very
shallow gradient for its central metallicity.

A correlation between the metallicity gradient and central age
was detected in many previous studies for galaxies 
in different environments (e.g. S{\'a}nchez-Bl{\'a}zquez, Gorgas \& Cardiel
2006a; S{\'a}nchez-Bl{\'a}zquez et al. 2007). Excluding the three galaxies with flat
metallicity gradients (see Sec. \ref{z_profile}), Fig. \ref{grdz1} (b)
shows that our galaxies 
follow similar trends as those found by S{\'a}nchez-Bl{\'a}zquez et al. (2007)  
for galaxies in higher density environments. 
Galaxies with young central ages tend to have steep metallicity
gradients while galaxies with uniform old stellar populations reveal
shallower metallicity gradients.
Since the galaxies with young central ages are those with the steepest
age gradients (Fig. \ref{grdage1}), we expect the gradient
of the total metallicity and the age gradient to be strongly
correlated. Fig. \ref{gzgage} shows that galaxies of steep 
positive age gradients also show steep negative metallicity
gradients. This implies that the young central stellar populations are also more
metal-rich than the old metal-poor stars in the outer regions of the
galaxies.

\begin{figure}
\centerline{\psfig{figure=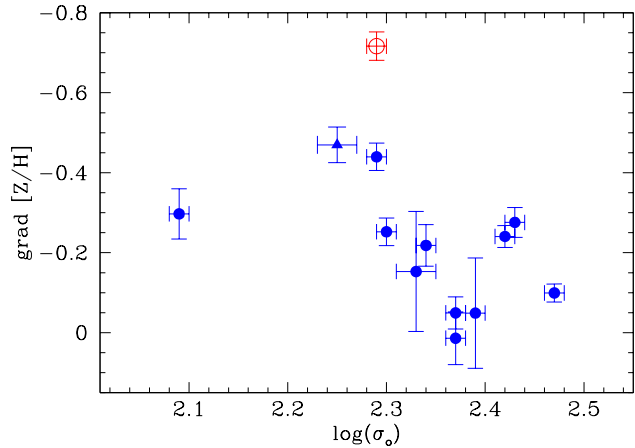,width=0.37\textwidth,angle=-90}}
\caption
{
The radial gradient of [Z/H] as a function of the central
velocity dispersion log($\sigma_o$). A strong anti-correlation is seen
for galaxies with log($\sigma_o$)$\ge2.2$. Symbols as in Fig. \ref{centr}.
}
\label{mass1}
\end{figure}

Fig. \ref{mass1} shows that galaxies of smaller mass (lower
log($\sigma_o$))  tend to reveal steeper metallicity gradients than
more massive galaxies (larger log($\sigma_o$)). 
A hint of a similar trend was found by S{\'a}nchez-Bl{\'a}zquez et
al. (2007) for galaxies of log($\sigma_o$) $\ge 2.2$ ($\sim 160$~km/s).
A change in the trend direction at this velocity dispersion, has been
reported in several studies (see S{\'a}nchez-Bl{\'a}zquez et al. 2007 and
references therein). Thus for galaxies with log($\sigma_o$) $\le 2.2$,
metallicity 
gradients appear to get shallower for {\it lower} mass (e.g. Forbes et al.
2005). We note that NGC 4240 (log($\sigma_o$)$=2.09$) is consistent
with this suggestion.

Examining the correlation of the metallicity gradient with galaxy
dynamical mass log($\sigma_o^2 r_e$) and the absolute
magnitude in the K-band (as other proxies of mass) gives
similar trends to that of Fig. \ref{mass1}.

\begin{figure*}
\centerline{\psfig{figure=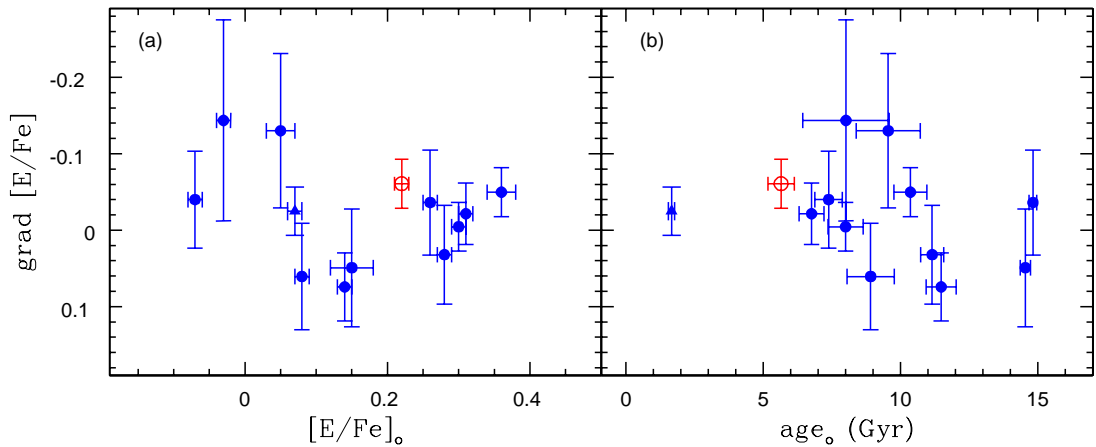,width=0.37\textwidth,angle=-90}}
\caption
{
The radial gradient of the $\alpha$-element abundance [E/Fe] versus a)
the central values [E/Fe]$_o$ and b) the central age. There is no
obvious correlation found in the two panels. Symbols as in
Fig. \ref{centr}.   
}
\label{grdefe1}
\end{figure*}

For a sample of early-type galaxies Ferreras \& Silk (2002) found a
correlation between [E/Fe] 
gradient and its central value. A steep negative 
gradient for low [E/Fe] galaxies changes to a steep positive gradient
for enhanced [E/Fe] galaxies. Using a simple model of star formation
and a standard prescription for the rates of supernovae Type II and Ia,
Ferreras \& Silk found that negative gradients imply inside-out
formation which is suggestive of a dissipative collapse formation. On 
the other hand, a positive gradient is a sign of outside-in formation
which may be a result of a past merger events which induce central
star formation in the merger remnant. Our isolated galaxies show
statistical insignificant [E/Fe] gradients on average. Furthermore,
Fig. \ref{grdefe1} shows that there is no obvious correlation between
[E/Fe] gradient and its central value or the central age.

\section{Discussion}

Several galaxies in our
isolated sample reveal a number of features such as tidal tails, dust,
shells, discy and boxy isophotes and highly rotating central discs
(Reda et al. 2004, 2005; Hau \& Forbes 2006). These structures indicate
recent merger/accretion events during the evolutionary
history of these galaxies. Measuring the age of the stellar
populations of the isolated galaxies reveals that several galaxies
have central young stars which also requires a recent gaseous
accretion or merger. 

If isolated galaxies formed purely by dissipative collapse we
would expect uniformly old stellar ages. We find both young
central ages and age gradients in some isolated
galaxies. Collapse models (e.g. Carlberg 1984;  Chiosi \& Carraro
2002; Kawata \& Gibson 2003) 
predict steeper metallicity gradients in more massive
galaxies.  For the bulk of our sample, with velocity dispersions
$\ge160$ km/s, we find the opposite behaviour with more massive
galaxies have shallower metallicity gradients. A collapse also
implies a largely inside-out formation with the inner galaxy
regions forming first which predicts a negative
[E/Fe] gradient (e.g. Ferreras \& Silk 2002). We find isolated galaxies to
reveal no overall [E/Fe] gradient.
Thus we conclude
that pure dissipative collapse can not explain our sample of
relatively massive isolated early-type galaxies.

The expectations for stellar population parameters in a
hierarchical merger scenario are more varied depending on the
gas fraction of the progenitors and their mass ratio 
(e.g. Kobayashi 2004). Gas-rich mergers are expected to induce central
star formation (e.g. Mihos \& Hernquist 1996; Springel 2000).
The gradients of age and metallicity are predicted to correlate
strongly with central young stars that are also metal-rich (Bekki \&
Shioya 1999). The metallicity gradient is also predicted to be
shallower for a merger remnant (e.g. Kobayashi 2004).

The age gradient for the majority of our isolated galaxies (11 out of
13) are found to range from flat to positive.
 These age gradients are also correlated with the central age which
 implies that a young burst dominates the luminosity in the central
 region where it takes place, and thus produces a steep gradient. As
 these young populations get older and fade, the observed gradient of
 the luminosity-weighted age becomes shallower. 
However, two galaxies (NGC 1045 and MCG-02-13-009) show significant
negative age gradients. The radial age profile of NGC 1045
reveals a stellar 
population of intermediate age at $r/r_e\sim0.2$ and outward, which
implies a secondary star formation event induced in the outer regions of
the galaxy. That suggests either a late gas accretion or  merger
with a gas-rich satellite galaxy (Kobayashi 2004). Such accretion may
form a star forming ring. Deep imaging of
this galaxy reveals strong boxy isophotes  
and extensive extra tidal light (Reda et al. 2004) which argues in
favour of a recent merger.  
While deep imaging is not available for
MCG-02-13-009, its negative age gradient and flat metallicity
gradient in the central region is predicted by the simulations of
Kobayashi (2004) to be the result of late gas accretion rather than a
major merger. 

Most of our isolated galaxies (8 out of 13) are found to have negative
metallicity gradients, as galaxies with central metal-rich stars
tend to have steeper metallicity gradients. 
The three galaxies NGC 1162, NGC 2865 and NGC 821 have both
significant age 
gradients as well as metallicity gradients, with the young central
stellar population also being more metal-rich than the old metal-poor
stars in the outer regions. 

NGC 1162 has an intermediate central age of 6.8 Gyrs suggesting a
secondary star formation event. A major merger of
two massive progenitors is expected to produce a more metal-rich
stellar population
and shallower metallicity gradient than observed. However, a
minor merger with a satellite of mass less than 0.2 times that of the 
primary galaxy is more likely to explain the intermediate central
metallicity and the gradient (Kobayashi 2004).

The galaxy NGC 2865 is the youngest (1.7 Gyrs) and most metal-rich
([Z/H]$_o=0.48$) galaxy of our sample. Its central velocity
dispersion of 178 km/s indicates an intermediate mass galaxy. 
Deep imaging revels a surrounding shell
structures (Hau, Carter \& Balcells 1999; Reda et
al. 2004) and kinematics reveal a kinematically decoupled core in the
central 4 arcsec or 6.4 kpc (Hau, Carter \& Balcells 1999; Hau \&
Forbes 2006). The kinematics and stellar population of
this galaxy suggest a gas-rich accretion or merger origin for the
shell and kinematics structures (Hau et al. 1999). 
However, the steep metallicity gradient of --0.47 prefers a gas-rich
accretion by a metal-rich parent galaxy. 

The galaxy NGC 821 has a central stellar population of 5.6 Gyrs old and
a high metallicity of [Z/H]$_o=0.4$, which indicates a secondary star
burst. 
Gas accretion or a minor merger would produce a lower central 
metallicity, while a major merger would have reduced the very steep
metallicity gradient of --0.72. In the detailed study of Proctor et
al. (2005), NGC 821 is speculated to consume its own gas to fuel a
secondary central burst of star formation.

The metallicity gradient of isolated galaxies are found to be
steeper for less massive galaxies. This result is consistent with a
major merger history for the massive galaxies (e.g. Bekki \& Shioya
1999). During the merger process, the high metallicity stars in the
centre of the galaxy are transported to the outer regions which
dilutes the metallicity gradient (Bekki \& Shioya 1999). On the
other hand, 
the induced star formation during the merger at the centre of the
merger remnant is expected to maintain the  metallicity gradient to
some extent. The final metallicity gradient of the merger remnant is
shallower than the progenitors and the newly formed central stars 
 will also produce relatively steep age gradients for these galaxies 
(Kobayashi 2004).
The gas-rich merger simulation of Bekki \& Shioya (1999) predicts a
one-to-one correlation between the age gradient and metallicity
gradient. However, our isolated galaxies show a correlation of slope
$0.6\pm0.1$. 

Both isolated galaxies, and those in HDEs, show similar age-mass
relations as more massive galaxies are older than smaller ones.
That suggests either the less massive
galaxies have had successive star formation events (bursts induced by
mergers) continuing 
until recent epochs, while more massive galaxies did not suffer such
events. Alternatively, all galaxies started forming their stars in a
single burst (similar to a dissipative collapse) at the same epoch
which stopped earlier in more massive galaxies, while less 
massive galaxies continue forming new stars for extended epochs.
Several mechanisms have been introduced to suppress the star
formation processes in more massive galaxies such as a central active
galactic nuclei (e.g. Springel, Di Matteo \& Hernquist 2005; De Lucia
et al. 2006). The dependence of the 
star formation efficiency on the circular velocity of the galaxy can
also delay the star formation in small galaxies relative to more massive
galaxies (e.g. De Lucia et al. 2004).

The old galaxies in our sample (NGC 2271, ESO153-G003, ESO218-G002 and
MCG-03-26-030) with central age $\ge$ 11 Gyrs show
flat age gradients which indicates that all stars are formed at same
time at $z\ge2$. They also tend to be the most massive galaxies.  
This can be a result of a single burst of star formation during a
rapid collapse of a single cloud (Larson 1974, 1975; Calberg 1984). 
The observed metallicity gradient of these galaxies ($<-0.28$) 
is much less than the predicted gradients of the dissipative collapse
models. The major merger scenario
can explain the shallow metallicity gradients of these galaxies 
but not their uniform old populations.
Another mechanism such as developing central active galactic nuclei
(e.g. Springel, Di Matteo \& Hernquist 2005; De Lucia et al. 2006) is
necessary to stop any further 
star formation after the merger. The low metallicity of ESO153-G003
([Z/H]$_o=0.19$) suggests progenitors of relatively small size for this
galaxy. 
The galaxy NGC 2271 has the highest metallicity among these galaxies
([Z/H]$_o=0.44$) with a flat metallicity gradient of. It also shows
elongated shape ($\epsilon=0.3$) and a solid body rotation  
with ($V/\sigma_o)^*=1.45\pm0.12$ (Hau \& Forbes 2006) which may
indicate amassive disk progenitor.  

The shallow metallicity gradients of the two galaxies NGC 682 and NGC
4240 indicate past major mergers. On the other hand, their uniform
luminosity weighted intermediate ages suggest a very small fraction of
induced star formation. 

The magnitude of the metallicity gradients are found to correlate to
the age gradient. 
The suggested mechanism causing these gradients is that each star
formation event will enrich the ISM, followed by radial inflow,
causing the subsequent stellar generations at the galaxy centre to be
more metal-rich, as well as younger. A natural 
result of this process is the observed age-metallicity relation. 
The dynamical effect of supernova feedback is expected to be
relatively weaker 
in more massive galaxies, consequently the star formation rate is more
efficient in these galaxies than in less massive ones.
The deep potential well of massive galaxies maintains their
gas long enough to perform more complete chemical processes to produce
higher metallicities (e.g. Arimoto \& Yoshii 1987; Edmunds 1990;
Matteucci 1994) resulting in the 
observed mass-metallicity relation. 

Independent of the galaxy mass, or the details of the merging process,
the observed correlation between the metallicity gradient and its
central value can be reproduced by the gas-rich merger simulations of
Bekki \& Shioya (1999).  The results of these models are also consistent
with the observed insignificant gradients of [E/Fe] within 1
effective radius of the isolated galaxies.

The parameter [E/Fe]$_o$ measures the abundance ratio of the
$\alpha$-element to 
the Fe-peak elements which are predicted to be released to the
interstellar medium by supernovae type II and Ia and on different time
scales. In 
that sense, [E/Fe]$_o$ is commonly used to quantify the duration of
star formation (Worthey, Faber \& Gonzalez 1992; Matteucci 1994;
Thomas, Greggio \& Bender 1999; Thomas et al. 2005). For the isolated 
galaxies as well as in HDEs, the lower [E/Fe]$_o$ of
the less massive galaxies points to 
extended star formation for these galaxies. However, the isolated
early-type galaxies tend  to have younger 
ages and lower [E/Fe]$_o$ for their central velocity dispersion than
their counterparts in HDEs. Furthermore, we also note that isolated
galaxies of intermediate ages 
tend to have lower [E/Fe]$_o$ than the galaxies in HDEs. This is
expected if the extended star 
formation is triggered with recent mergers which have stopped in HDEs at
higher redshifts but continue to recent
epochs for galaxies in low density environments such as our isolated
galaxies.

\section{Conclusions}

Isolated galaxies show a wide range of ages, [Z/H] and [E/Fe]
for their central stellar populations. More massive galaxies tend to
be older, more metal-rich and more $\alpha$-element enhanced than less
massive galaxies in the same sense as galaxies in high density
environments. Although isolated galaxies tend to be slightly
younger, more metal-rich and lower [E/Fe] for their mass.

The majority of the isolated galaxies in our sample show insignificant
gradients of both age and [E/Fe]. Although a correlation between the
age gradient and central age indicates the tendency of the newly formed
stars to locate closer to the galaxy centre.

The metallicity gradient ranges from very steep to
flat. Metallicity gradients are found to correlate with parameters of
the central stellar populations such as central metallicity, age and
velocity dispersions. Metallicity gradients also show a remarkable
correlation with the age gradients.

The formation scenario of a single dissipative collapse cloud can not
explain the spatial distribution of the stellar population and
kinematic properties found for isolated galaxies. Mergers at
different redshifts of progenitors of different mass ratios and gas
fractions are needed to reproduce the observed properties of the
galaxies.

\section{References}

Arimoto N., Yoshii Y., 1987, A\&A, 173, 23 \\
Bekki K., Shioya Y., 1999, ApJ, 513, 108 \\
Bernardi et al., 2003, AJ, 125, 1882 \\
Bernardi M., Nichol R.~C., Sheth R.~K., Miller C.~J., Brinkmann J.,
    2006, AJ, 131, 1288 \\ 
Brough et al., 2006, supmitted to MNRAS \\
Carlberg R.G., 1984, Apj, 286, 403 \\
Chiosi C., Carraro G., 2002, MNRAS, 335, 335  \\
Colbert J.~W., Mulchaey J.~S., Zabludoff A.~I., 2001, AJ, 121, 808 \\ 
Collobert M., Sarzi M., Davies R.~L., Kuntschner H., Colless M., 2006,
     MNRAS, 370, 1213 \\ 
De Lucia G., Kauffmann G., White S.~D.~M., 2004, MNRAS, 349, 1101 \\ 
De Lucia G., Springel V., White S.~D.~M., Croton D., Kauffmann G.,
     2006, MNRAS, 366, 499 \\ 
Denicol{\'o} G., Terlevich R., Terlevich E., Forbes D.~A., Terlevich
     A., 2005b, MNRAS, 358, 813  \\
Denicol{\'o} G., Terlevich R., Terlevich E., Forbes D.~A., Terlevich
     A., Carrasco L., 2005a, MNRAS, 356, 1440 \\ 
Edmunds M. G., 1990, MNRAS, 246, 678 \\
Ferreras I., Silk J., 2002, MNRAS, 336, 1181 \\
Forbes D.~A., S{\'a}nchez-Bl{\'a}zquez P., Proctor R., 2005, MNRAS,
   361, L6 \\ 
Hau G.~K.~T., Carter D., Balcells M., 1999, MNRAS, 306, 437 \\
Hau G.~K.~T., Forbes D.~A., 2006, MNRAS, 371, 633 \\
Jarrett T. H., Chester T., Cutri R., Schneider S. E., Huchra J. P.,
    2003, AJ, 125, 525 \\ 
Kawata D., Gibson B.~K., 2003, MNRAS, 346, 135 \\
Kelson D. D., Illingworth G. D., Franx M., van Dokkum P.G.,
    2006, astro-ph/0606642 \\
Kobayashi C., 2004, MNRAS, 347, 740 \\
Kobayashi C., Arimoto N., 1999, ApJ, 527, 573 \\
Kuntschner H., Lucey J.~R., Smith R.~J., Hudson M.~J., Davies R.~L.,
    2001, MNRAS, 323, 615 \\ 
Kuntschner H., Smith R.~J., Colless M., Davies R.~L., Kaldare R.,
    Vazdekis A., 2002, MNRAS, 337, 172 \\ 
Larson R.~B., 1974, MNRAS, 166, 585 \\
Matteucci F, 1994, A\&A, 288, 57 \\
Mehlert D., Thomas D., Saglia R. P., Bender R., Wegner G., 2003, A\&A,
    407, 423 \\
Mihos J. C., Hernquist L., 1996, ApJ, 464, 641 \\
Proctor R.~N., Forbes D.~A., Beasley M.~A., 2004, MNRAS, 355, 1327 
Proctor R. N., Forbes D. A., Forestell A., Gebhardt K., 2005,
   MNRAS, 362, 857\\
Proctor R.~N., Forbes D.~A., Hau G.~K.~T., Beasley M.~A., De
   Silva G.~M., Contreras R., Terlevich A.~I., 2004, MNRAS, 349,
   1381 \\ 
Proctor R. N., Sansom A. E., 2002, MNRAS, 333, 517 \\
Prugniel P., Simien F., 1997, A\&A, 321, 111 \\
Reda F.~M., Forbes D.~A., Beasley M.~A., O'Sullivan E.~J., Goudfrooij P., 
   2004, MNRAS, 354, 851 \\
Reda F.~M., Forbes D.~A., Hau G.~K.~T., 2005, MNRAS, 360, 693 \\
S{\'a}nchez-Bl{\'a}zquez P., Forbes D.~A., Strader J., Brodie J., Proctor R.,
    2007, submitted to MNRAS \\
S{\'a}nchez-Bl{\'a}zquez P. Gorgas J., Cardiel N., 2006a, A\&A, 457, 823\\
S{\'a}nchez-Bl{\'a}zquez P., Gorgas J., Cardiel N., Gonzalez J. J., 2006b,
    A\&A, 457, 809 \\
Springel V., 2000, MNRAS, 312, 859 \\
Springel V., Di Matteo T, Hernquist L., 2005, ApJ, 620, L79 \\
Terlevich A. I., Forbes D. A., 2002, MNRAS, 330, 547 \\
Thomas D., Greggio L., Bender R., 1999, MNRAS, 302, 537 \\
Thomas D., Maraston C, Bender R., 2003, MNRAS, 339, 897 \\
Thomas D., Maraston C, Bender R., de Oliveira C. M., 2005, ApJ, 621,
    673 \\ 
Thomas D., Maraston C., Korn A., 2004, MNRAS, 351, L19 (TMK04) \\
Trager S.~C., Worthey G., Faber S.~M., Burstein D., Gonzalez J.~J., 1998, 
    ApJS, 116, 1 \\
Vazdekis A., 1999, ApJ, 513, 224 \\
White S. D. M., 1980, MNRAS, 191,1 \\
Worthey G., Faber S. M., Gonzalez J. J., 1992, ApJ, 398, 69\\
Worthey G., Ottaviani D.~L., 1997, ApJS, 111, 377 \\

\end{document}